\documentclass{article}
\usepackage{amsmath,mathrsfs}
\usepackage{graphicx,subfigure}
\usepackage{multirow,booktabs}
\usepackage[numbers,sort&compress]{natbib}

\def\abstract#1{\vskip 7mm 
        \begin{center}{\large Abstract}\par \smallskip
                \begin{minipage}[c]{12cm}
                        \small #1
                \end{minipage}
        \end{center}
}
\def\title#1{\begin{center}{\Large\bf #1}\end{center}}
\def\author#1{\vskip 5mm \begin{center}{#1}\end{center}}
\def\address#1{\begin{center}{\it #1}\end{center}}
\makeatletter
\@ifundefined{lesssim}{}{}
\@ifundefined{gtrsim}{}{}
\def\vereq#1#2{\lower3pt\vbox{\baselineskip1.5pt \lineskip1.5pt
\ialign{$\m@th#1\hfill##\hfil$\crcr#2\crcr\sim\crcr}}}
\makeatother

\begin{document}

\title{%
  The Uplifting of AdS Type to Quintessence Like Potential Induced by Frozen Large Scale Lorentz Violation
}
\author{ Hanyu Zhai$^a$, Jiayin Shen$^a$ and Xun Xue$^{a,b,}$\footnote{Corresponding author:xxue@phy.ecnu.edu.cn}}
\address{ $^a$Department of Physics, East China Normal University, Shanghai 200241, China}
\address{$^b$Center for Theoretical Physics, Xinjiang University, Urumqi 830046, China}

\abstract{
 The quintessence-like potential of vacuum energy can meet the requirement from both quantum gravity and the accelerating expansion of the  universe. The anti-de Sitter vacuum in string theory has to be lifted to the meta-stable de Sitter vacuum with positive vacuum energy density to explain the accelerating expansion of the universe. Based on the possible large scale Lorentz violation, we define an effective cosmological constant which depends not only on the bare cosmological constant but also on the Lorentz violation effect. We find the evolution of the effective cosmological constant exhibits the behavior of quintessence potential when the bare cosmological constant is from string landscape in contrary to the existence of local minimum during evolution while the bare cosmological constant is supplied by the swampland. The critical value of bare cosmological constant is approximately zero for the behavior transition. The frozen large scale Lorentz violation can uplift the AdS vacua to an effective quintessence-like one in this sense. 
}

\section{Introduction}
It is widely accepted that all the four kinds of basic interactions will be unified into quantum gravity at the Planck energy scale and the universe is dominated by quantum gravity at the beginning of creation. The theory describing quantum gravity is far from completion. String theory is one of the efforts. There are many different approaches, such as loop quantum gravity, Ho$\check{r}$ava-Lifshitz gravity, non-commutative geometry etc. Many of them share the common feather of explicit Lorentz violation(LV) and part of them are non-local theories and implicit Lorentz violated\cite{carroll2001noncommutative,alfaro2004alternative}. 

The observable universe is believed to be expanded from a very tiny part of spacetime in the era of quantum gravity domination by inflation. The scale of area with Lorentz violation is expanded so fast as to exceed the horizon in a very short time. Different parts of the region interacting via quantum gravity lose interaction at the instance exceeding out of horizon and Lorentz violation of the small region before inflation is frozen in the large scale after inflation in this way. The scale may re-enter the horizon during the period of normal expansion depending on details of quantum gravity and inflation physics\cite{Shen:2018elj}. 

There are also possible signs of large scale Lorentz violation in the cosmic observations, e.g. the anisotropies in the low-l multi-pole expansion of CMB power spectrum. The normals of quadra-pole and octopole are not coincided with the direction of dipole which indicates the boost transformation from CMB static frame to the peculiar motion frame is not a simple Lorentz boost\cite{Perivolaropoulos2014}. The significant tension between values of Hubble constant directly measured by the distance ladder method and the measurement by $\Lambda$CDM model from CMB observation is possibly another sign of physics beyond the standard cosmological model based upon local Lorentz invariant gravitation theory, the $\Lambda$CDM model\cite{Riess:2019cxk}. 

General relativity is a very successful gravitation theory at least at small scale. When general relativity is applied to the larger scale, there appear some deviations from its predictions which are usually not recognized as deviations but as effects caused by unknown energy-momentum distribution named as dark matter and dark energy in turn. One has to take dark energy into account when dealing with the cosmic scale physics. Due to the possible frozen Lorentz violation at large scale, the effective theory of gravity at the cosmic scale has to take Lorentz violation into account and it will be inevitably a non-Einstein gravitation theory for the theory of general relativity is local Lorentz invariant. Actually the $\Lambda$CDM model is very successful at explaining most cosmological phenomenon by adding cosmological constant term into Einstein theory of gravity,
\begin{equation}\label{LCDMaction}
{{S}_{E}}=\frac{1}{16\pi G}\int{{{d}^{4}}x}\sqrt{-g}\left( R-2\Lambda  \right)
\end{equation} 
or
\begin{equation}\label{LCDMequation}
{{R}_{\mu \nu }}-\frac{1}{2}R{{g}_{\mu \nu }}+\Lambda {{g}_{\mu \nu }}=8\pi G{{\left( {{T}_{M}} \right)}_{\mu \nu }} \, ,
\end{equation}
where $\Lambda$ is the cosmological constant which can also be regarded as the vacuum energy density. However, there is a big puzzle about cosmological constant when it is regarded as the vacuum energy density for the theoretical prediction on its value is of 54 to 112 order of 10 higher than the value fitted by observation. One has to take fine tuning as a solution which lowers the confidence level of the model so as to be hard accepted\cite{martin2012everything}. The idea of large scale Lorentz violation is first proposed in 2015\cite{wu2015effective}. The framework of constructing large scale effective gravity with LV is build up as a modified gravity model with gauge principle via the equivalence principle by utilizing the constrain dynamics and the very special relativity(VSR) symmetry\cite{wu2015effective,wu2016sim,yang2017VSR,wei2017E2}. 

On the other hand, the low energy theory of physics describing the Nature is believed to be an effective quantum field theory derived from a fundamental quantum gravity theory which is far from well understood and has many possible candidates. Although string theory can supply many possible links between the fundamental unified theory and the low energy effective theory, it has an obvious shortage that there are too many possible vacua, of the order of $10^{500}$, when compactifying the extra six dimension to get a four dimensional effective low energy theory. The vacuum energy of different cacua constituent a complex landscape\cite{Vafa:2005ui}. Anti-de Sitter type of vacua from flux compactification of string are generic and supersymmetry preserving. To account for the late time accelerating expansion of the universe, one need to lift the negative anti-de Sitter vacuum energy to a meta-stable positive de Sitter one with some unnatural techniques in string theory\cite{Kachru:2003aw,Cicoli:2008va}. 
However the low energy effective field theories built on these meta-stable de Sitter vacuum do not have a UV completion hence are classified as belonging to the swampland in contrary to the lanscape because they are not supposed to be consistent with quantum gravity are not able to derived from landscape\cite{Vafa:2005ui,Brennan:2017rbf}. 

To explain the accelerating expansion by $\Lambda$CDM model one seems to need a stable or meta-stable dS vacua to supply a minimum of the potential energy density as the cosmological constant. Unfortunately, the tension between swampland condition and inflation excludes the dS type of potential as favorable one leaving the only allowed potentials with quintessence and AdS types\cite{Agrawal:2018own}. Can the universe with an AdS vacuum expand accelerating? Is the quintessence a fundamental canonical scalar field or the quintessence type of potential emerged from other fundamental mechanism? We try to investigate the problem in a different view of point within the framework of effective gravity with large scale LV following the pioneer papers\cite{wu2015effective,wu2016sim,yang2017VSR,wei2017E2}.  

\section{Gravity with Large Scale Lorentz Violation}
Take $SIM(2)$ symmetry, the proper subgroup of Lorentz group, as an example in the language of tetrad field ${h_a}^{\mu}$. One can build local $SIM(2)$ gravity with gauge principle by constraining the gauge potential, the Lorentz connection, takes value only on $sim(2)$ algebra. The action for a $sim(2)$ gravity can be taken as
\begin{equation}\label{sim2action}
{{S}_{sim2}}\text{=}\frac{1}{16\pi G}\int{{{d}^{4}}x}h\left( {{R}^{ab}}_{ab}+{{\lambda }_{1}}^{\mu }\left( {{A}^{10}}_{\mu }-{{A}^{31}}_{\mu } \right)+{{\lambda }_{2}}^{\mu }\left( {{A}^{20}}_{\mu }+{{A}^{23}}_{\mu } \right) \right)
\end{equation} 
where the Lagrange multipliers are used to constrain the Lorentz connection taking values on $sim(2)$ algebra and ${h=\det {h^a}_\mu}$. The Lagrange-multipliers terms contribute an effective angular momentum distribution $C_{Meff}$ such that
\begin{equation}\label{effangularmomentum}
{{D}_{\nu }}\left( h\left( {{h}_{a}}^{\nu }{{h}_{b}}^{\mu }-{{h}_{a}}^{\mu }{{h}_{b}}^{\nu } \right) \right)=16\pi G{{\left( {{C}_{M}}\text{+}{{C}_{M}}_{eff} \right)}_{ab}}^{\mu } 
\end{equation}
where $C_M$ is the angular momentum distribution of the matter source. This equation leads to the torsion free condition in general relativity in the case of $C_M=0$ where $C_{Meff}=0$ always holds for the local Lorentz invariance. On the contrary, Lorentz violation leads to non-trivial distribution of torsion tensor even in the scalar matter source case where $C_M=0$. 
The variation respect to tetrad fields $\dfrac{\delta S_{sim2}}{\delta {{h}^{a}}_{u}}$ gives the equations of motion for tetrad fields,
\begin{equation}\label{sim2eisteineq}
{{R}^{a}}_{b}-\frac{1}{2}R{{\delta }^{a}}_{b}=\frac{8\pi G}{{{c}^{4}}}{{\left( {{T}_{M}} \right)}^{a}}_{b} \; ,
\end{equation}
where ${\left(T_M\right)^a}_b$ is the energy-momentum tensor of matter source. It should be noted that the connection hides in ${{R}^{a}}_{b}$ and $R$ is no longer the torsion free Levi-Civita one for the reason discussed in \eqref{effangularmomentum}. Generally, the connection can be decomposed into torsion-less part and contortion
\begin{equation}\label{decomposion}
{{A}^{a}}_{bc}={{\tilde{A}}^{a}}_{\; \;bc}+{{K}^{a}}_{bc}\, ,
\end{equation}
where $\widetilde{A}_{\; \;bc}^{a}$ is the Levi-Civita connection and ${{K}^{a}}_{bc}$ is the contortion part.\cite{Aldrovandi:2013wha}. 
The curvature can be decomposed into three parts with the help of decomposition of spin connection as
\begin{equation}\label{Rdecomposion}
{{R}^{mn}}_{ab}={{\tilde{R}^{mn}}}_{\;\;\;\;\;\;ab}+{R_{K}}{{^{mn}}_{ab}}+{{R}_{CK}}{{^{mn}}_{ab}} \; ,
\end{equation}
where ${{\tilde{R}^{mn}}}_{\;\;\;\;\;\;ab}$ and ${{R}_{K}}{{^{mn}}_{ab}}$ are
the curvatures composed of torsion-free connection and contortion respectively,
while ${{R}_{CK}}{{^{mn}}_{ab}}$ contains cross terms of them. In the geometric unit $\frac{8\pi G}{{{c}^{4}}}=1$, we can rewrite
\eqref{G=T} as
\begin{equation}\label{darkpartner}
{{\tilde{R}}^{a}}_{\;\;b}-\frac{1}{2}{{\delta }^{a}}_{b}\tilde{R}={{\left( T_{eff}+{{T}_{M}} \right)}^{a}}_{b} 
\; ,
\end{equation}

where ${\tilde{R}}_{c}^{\; \;a}$ and $\tilde{R}$ are generated by torsion free Levi-Civita connection ${{\tilde{A}}^{a}}_{\; \;bc}$ and ${T}_{eff}$ collects all the terms involving contortion ${{K}^{a}}_{bc}$,
\begin{equation}\label{Teff1}
{{\left( T_{eff} \right)}^{a}}_{b} =  \dfrac{1}{2}{{\delta }^{a}}_{b}\left( {{R}_{K}}+{{R}_{CK}} \right)-\left( {{R}_{K}}{{^{a}}_{b}}+{{R}_{CK}}{{^{a}}_{b}} \right)  \; .
\end{equation}

The effective energy-momentum tensor $T_{eff}$ contributes
to the gravitation in addition to matter contribution $T_{M}$
and appears as the dark partner of the matter distribution
The non-trivial effective contribution to the energy-momentum distribution by contortion is expected to be responsible for the dark partner of the matter,
The Bianchi identitity guarantees the conservation of ${T}_{eff}$ respected to the Levi-Civita connection ${{\tilde{A}}^{a}}_{\; \;bc}$.

Based upon our analysis on possible frozen large scale Lorentz violation, the cosmological observable actually are not Lorentz covariant but still $SO(3)$ covariant. Simply restricting the components of Lorentz gauge field ${A^a}_{b\mu}$ nontrivial only on $SO(3)$ generators in the construction of a gravitation theory with Lorentz boost violation would result in a degenerated dynamics. The reason is that boost transformation is not prohibited at the large scale actually for only the Lorentz boost transformation is violated. There are discussions on the modification of Lorentz algebra at quantum level by Hopf algebra or deformed Poincare algebra such as the  $\kappa$-Poincare etc. as well as other quantum gravity model like Horava-Lifshitz gravity in which the Lorentz boost is automatically violated. 

Observing that the Lorentz gauge potentials transform as
\begin{equation}\label{transA}
{{A'}^{a}}_{b\mu }={{\Lambda }^{a}}_{c}\left( x \right){{A}^{c}}_{d\mu }{{\Lambda }_{b}}^{d}\left( x \right)+{{\Lambda }^{a}}_{c}\left( x \right){{\partial }_{\mu }}{{\Lambda }_{b}}^{c}\left( x \right)
\end{equation}
under local Lorentz transformation $\Lambda(x)$, it is obvious that ${A'^0}_{i\mu} ={{\Lambda }_{i}}^{j}\left( x \right){{A}^{0}}_{j\mu }$ for a rotation transformation ${\Lambda\in SO(3)}$. Hence the restriction to the Lorentz gauge potentials can be proposed as
\begin{equation}\label{constraint}
\left({A^{0}}_{1\mu}\right)^2+\left({A^{0}}_{2\mu}\right)^2+\left({A^{0}}_{3\mu}\right)^2=\left(f_\mu (x)\right)^2
\end{equation}
where $f_\mu(x)$ can be regarded as a measurement of the magnitude of the boost violation in some sense, and it is invariant under a local $SO(3)$ gauge transformation on tetrad fields ${h_a}^{\mu}$ respected to the tetrad indices but frame dependent. 

The action of effective gravity with large scale Lorentz violation can then be given by, 
\begin{equation}\label{action}
S\text{=}\frac{{{c}^{4}}}{16\pi G}\int{{{d}^{4}}x}h\left( R-2{{\Lambda }_{0}}+{{\lambda }^{u}}\left( {{\left( {{A}^{0}}_{1u} \right)}^{2}}+{{\left( {{A}^{0}}_{2u} \right)}^{2}}+{{\left( {{A}^{0}}_{3u} \right)}^{2}}-f_{u}^{2} \right) \right)
\end{equation}
where the repeated superscript and the subscript $\mu$ of $\lambda^\mu$ and ${A^0}_{i\mu}$ respectively mean summation and ${\Lambda }_{0}$ is the bare cosmological constant given by the vacuum energy density. 

The variation respect to tetrad fields $\dfrac{\delta S}{\delta {{h}^{a}}_{u}}$ gives the equations of motion for tetrad fields,
\begin{equation}\label{G=T}
{{G}^{a}}_{b}\equiv {{R}^{a}}_{b}-\frac{1}{2}R{{\delta }^{a}}_{b}\text{+}{{\Lambda }_{0}}{{\delta }^{a}}_{b}=\frac{8\pi G}{{{c}^{4}}}{{\left( {{T}_{M}} \right)}^{a}}_{b} \; ,
\end{equation}
where ${\left(T_M\right)^a}_b$ is the energy-momentum tensor of all the matter source both luminous and dark one. As discussed in eq.\eqref{sim2eisteineq},  ${{R}^{a}}_{b}$ and $R$ are composed of connection with torsion. We can rewrite
eq.\eqref{G=T} in the form of eq.\eqref{darkpartner} with curvature composed only of torsion-less Levi-Civita connection and the contortion part contributes effectively as the dark partner of the matter distribution with
\begin{equation}\label{Teff}
{{\left( T_{eff} \right)}_{c}}^{a} = \left( \dfrac{1}{2}{{\delta }_{c}}^{a}\left( {{R}_{K}}+{{R}_{CK}} \right)-\left( {{R}_{K}}{{_{c}}^{a}}+{{R}_{CK}}{{_{c}}^{a}} \right) \right)-{{\Lambda }_{0}}{{\delta }_{c}}^{a} \; .
\end{equation}
At the cosmic scale, the effective energy-momentum tensor $T_{eff}$ is expected to contribute to the dark energy effectively and responsible for the accelerating expansion.
In a given basis of tetrad ${h_{a}}^{\mu}$, the equations of motion for connection can be written down explicitly
\begin{equation}\label{connection}
{\mathcal{D}_{\nu }}\left( h{{h}_{0}}^{[\nu }{{h}_{i}}^{\mu ]} \right)\text{+}\frac{1}{2}\lambda^\mu h{{A}^{0}}_{i\mu }\text{=0} 
\end{equation}
where $i=1,2,3$ and the repeated superscript and subscript $\mu$ does not mean summation here for it is a result of variation of the square power of  ${A^0}_{i\mu}$ in eq.\eqref{action} and 
\begin{equation}\label{connection1}
{\mathcal{D}_{\nu }}\left( h{{h}_{i}}^{[\nu }{{h}_{j}}^{\mu ]} \right)\text{=}0	
\end{equation}
for the $i,j$ indices combination. 
With the decomposion ${{A^a}_{b\mu}}$ into Levi-Civita connection ${\widetilde A^a}_{\;\;b\mu}$ and contortion ${K^a}_{b\mu}$ in eq. \eqref{decomposion},
Eqs.\eqref{connection1} can be expressed in detail as
\begin{equation}\label{K}
\begin{split}
{K^0}_{12}={K^0}_{21},\;{K^1}_{23}=0,\;{K^2}_{12}=-{K^0}_{10},\;{K^3}_{13}=-{K^0}_{10},\\
{K^0}_{23}={K^0}_{32},\;{K^2}_{31}=0,\;{K^3}_{23}=-{K^0}_{20},\;{K^1}_{21}=-{K^0}_{20},\\
{K^0}_{31}={K^0}_{13},\;{K^3}_{12}=0,\;{K^1}_{31}=-{K^0}_{30},\;{K^2}_{32}=-{K^0}_{30}.
\end{split}
\end{equation}
and eqs.\eqref{connection} as
\begin{equation}\label{K0}
\begin{split}
&2{K^0}_{10}{h_0}^\mu+\left({K^0}_{22}+{K^0}_{33}\right){h_1}^\mu-\left({K^1}_{20}+{K^0}_{21}\right){h_2}^\mu +\left({K^3}_{10}-{K^0}_{31}\right){h_3}^\mu\\
&+\lambda^\mu\left({A^0}_{10}{h^0}_\mu +{A^0}_{11}{h^1}_\mu+{A^0}_{12}{h^2}_\mu+{A^0}_{13}{h^3}_\mu\right)=0 \\
&2{K^0}_{20}{h_0}^\mu+\left({K^1}_{20}-{K^0}_{12}\right){h_1}^\mu+\left({K^0}_{11}+{K^0}_{33}\right){h_2}^\mu-\left({K^2}_{30}+{K^0}_{32}\right){h_3}^\mu\\
&+\lambda^\mu\left({A^0}_{20}{h^0}_\mu +{A^0}_{21}{h^1}_\mu+{A^0}_{22}{h^2}_\mu+{A^0}_{23}{h^3}_\mu\right)=0\\
&2{K^0}_{30}{h_0}^\mu-\left({K^3}_{10}+{K^0}_{13}\right){h_1}^\mu+\left({K^2}_{30}-{K^0}_{23}\right){h_2}^\mu+\left({K^0}_{11}+{K^0}_{22}\right){h_3}^\mu\\
&+\lambda^\mu\left({A^0}_{30}{h^0}_\mu+{A^0}_{31}{h^1}_\mu+{A^0}_{32}{h^2}_\mu+{A^0}_{33}{h^3}_\mu\right)=0
\end{split}
\end{equation}
Actually the tetrad fields ${h_a}^{\mu}$ satisfy eqs. \eqref{K}, \eqref{K0} and eq. \eqref{G=T} as well as the constrains conditions\eqref{constraint} simultaneously together with the contortion ${K^a}_{bc}$ and the $\lambda$ multipliers. To solve the simultaneous equations, one can employ the cosmological principle which holds in the CMB static reference frame. The symmetry requirement by cosmological principle gives that the metric for the universe must have the form of Robertron-Walker space-time metric,  
\begin{equation}\label{FRW}
{\mathrm ds^2=\mathrm dt^2-a(t)^2\left(\dfrac{\mathrm dr^2}{1-kr^2}+r^2\mathrm d\theta^2+r^2\sin^2\theta\,\mathrm d\varphi^2\right)}\, .
\end{equation}
The instantaneous co-moving tetrad basis can be read out directly, 
\begin{equation}\label{guage}
h^0=\mathrm dt,\;
h^1=\frac{a(t)}{\sqrt{1-kr^2}}\,\mathrm dr,\;
h^2=a(t)r\,\mathrm d\theta,\;
h^3=a(t)r\sin\theta\,\mathrm d\varphi\; .
\end{equation}

The cosmic media should be a perfect fluid by the analysis of cosmological principle. The energy momentum $T_M$ has the character of a perfect fluid and can be described by the energy density $\rho$ and pressure $p$, i.e. with the diagonal form of ${{\left( {{T}_{M}} \right)}^{a}}_{\text{b}}=diag\left( \rho ,-p,-p,-p \right)$. The perfect fluid energy momentum tensor $T_M$ in eq.\eqref{G=T} requires ${{G^a}_{b}=0,\;\forall a\not=b}$. One can get the conclusion that the only possible non-zero independent components of contortion ${{K}^{a}}_{bc}$ are ${{K^0}_{11},{K^0}_{22},{K^0}_{33}}$ while the others can all be determined to be zero.

The cosmological principle also requires all cosmic physical quantities depend only on cosmic time $t$, and hence
\begin{equation}\label{K0ii}
{K^0}_{11}={K^0}_{22}={K^0}_{33}=\mathscr K(t)
\end{equation}
are the solutions required. The dependence between $\mathscr K(t)$ and $f_\mu(x)$ can be derived as
\begin{equation}\label{fmu}
\left(f_t,f_r,f_\theta,f_\varphi\right)=\left(a(t)\mathscr K(t)+\dot a(t)\right)\cdot\left(0, \frac{1}{\sqrt{1-kr^2}}, r, r\sin\theta\right)
\end{equation}
It is a reasonable result that the only one remaining of the four degrees of freedom for $f_\mu(x)$ is expressed by $\mathscr K(t)$ while the other three  are used to fix the reference frame ${h_a}^{\mu}$.

\section{Late time expansion of the universe in landscape and swampland}
Denote the contribution to the energy momentum tensor ${{\left( T_{eff} \right)}_{c}}^{a}$ of eq.\eqref{Teff} in the co-moving frame of Robertson-Walker universe from dark partner of a perfect fluid energy momentum tensor ${{\left( {{T}_{M}} \right)}^{a}}_{\text{b}}=Diag\left( \rho ,-p,-p,-p \right)$ as 
\begin{equation}\label{TK}
{{T}_{K }}{{^{a}}_{c}}\equiv \left( T_{eff} \right){{^{a}}_{c}}=Diag({{\rho }_{K}},-{{p}_{K }},-{{p}_{K }},-{{p}_{K }})
\end{equation}
We can get
\begin{equation}\label{rhoK}
{{\rho }_{K }}\text{=}-\left( 3{{\mathscr K}^{2}}+6\mathscr K\dfrac{{\dot{a}}}{a}-{{\Lambda }_{0}} \right)
\end{equation}
and 
\begin{equation}\label{pK}
{{p}_{K}}= {{\mathscr K}^{2}}+4\mathscr K\dfrac{{\dot{a}}}{a}+2\dot{\mathscr K}-{{\Lambda }_{0}} \; .
\end{equation}

Since astronomical observations reveal that the space of the universe is flat, i.e. $k=0$ in the Robertson-Walker metric ansatz\eqref{FRW} , we then get the large scale Lorentz violation modified Friedmann Equation,
\begin{equation}\label{mfeq1}
{{\left( \dfrac{{\dot{a}}}{a} \right)}^{2}}\text{=}\dfrac{\rho \text{+}{{\rho }_{K}}}{3}\text{=}\dfrac{\rho \text{+}{{\Lambda }_{0}}}{3}-{{\mathscr K}^{2}}-2\mathscr K\frac{{\dot{a}}}{a}
\end{equation}
and
\begin{equation}\label{mfeq2}
\ddot{a}=-\dfrac{a}{2}\left( p+{{p}_{K }}+\dfrac{\rho +{{\rho }_{K }}}{3} \right)\text{=}-\dfrac{a}{2}\left( p+\frac{\rho }{3} \right)\text{+}\dfrac{1}{3}a{{\Lambda }_{0}}-\dfrac{d}{dt}\left( a\mathscr K \right)
\end{equation}

From the modified Friedmann equation\eqref{mfeq2}, the condition for accelerating expansion of the universe can be easily obtained as ${\dfrac{a}{2}\left(p+\dfrac{\rho}{3}-\dfrac{2}{3}\Lambda_0 \right)+\dfrac{\mathrm d}{\mathrm dt}\left(a\mathscr K\right)<0}$ . As discussed in \cite{Shen:2018elj}, the prediction of Lorentz violation parameters $f_\mu(x)$ needs quantum gravity and the inflation model in some detail rather than the present model. However, it would be suggestive to seek some phenomenological approximations about $\mathscr K(t)$ instead of constructing the quantum gravity theory and inflation model in detail at first step. It is inspiring to compare the Friedmann equations of $\Lambda$CDM model,
\begin{equation}\label{feq1}
{{\left( \dfrac{{\dot{a}}}{a} \right)}^{2}}\text{=}\dfrac{\rho \text{+}{{\rho }_{\Lambda}}}{3}\text{=}\dfrac{\rho \text{+}{{\Lambda }}}{3}
\end{equation}
and
\begin{equation}\label{feq2}
\ddot{a}=-\dfrac{a}{2}\left( p+{{p}_{\Lambda }}+\dfrac{\rho +{{\rho }_{\Lambda }}}{3} \right)\text{=}-\dfrac{a}{2}\left( p+\frac{\rho }{3} \right)\text{+}\dfrac{1}{3}a{{\Lambda }}\; ,
\end{equation}
with the modified one \eqref{mfeq1} and \eqref{mfeq2}, where ${{\rho }_{\Lambda }}\text{=}\Lambda $, ${{p}_{\Lambda}}\text{=}-\Lambda $, $\dfrac{\Lambda}{3{H_0}^2}\approx 0.713775$ and $H_0$ is the present Hubble constant. It should be noted that $\Lambda$ in eq.\eqref{feq1} and \eqref{feq2} is an observation input or a theoretical input by a more fundamental theory. Once the equation of state(EoS), $p=w\rho$, of the cosmic media is given, one can make prediction on the evolution of cosmic observables $a$, $p$ and $\rho$ etc.. In the modified case of \eqref{mfeq1} and \eqref{mfeq2}, one need one more input, $\mathscr K$ which should be given by more fundamental theory, quantum gravity, or also as an observational input phenomenologically just like $\Lambda$ in eq.\eqref{feq1} and \eqref{feq2} of $\Lambda$CDM model. Since $\Lambda$CDM model works well phenomenologically, we can employ it to fix the evolution of $\mathscr K$ phenomenologically and approximately. 

Eq.\eqref{mfeq1} can be rewritten into
\begin{equation}\label{mht}
H ^2 (t) =\dfrac{\rho \text{+}{{\Lambda }_{0}}}{3}-{{\mathscr K}^{2}}-2\mathscr K H(t)
\end{equation}
the evolution equation for Hubble constant $H(t)=\dfrac{\dot a(t)}{a(t)}$. So can be eq.\eqref{feq1} as
\begin{equation}\label{ht}
H ^2 (t)=\dfrac{\rho \text{+}{{\Lambda }}}{3}
\end{equation}
with the solution
\begin{equation}\label{sht}
H(t) \text{=}\sqrt{\dfrac{\rho \text{+}{{\Lambda }}}{3}}\; .
\end{equation}
The solutions of eq.\eqref{mht} can then be chosen
\begin{equation}\label{smht1}
H(t) \text{=}\sqrt{\dfrac{\rho \text{+}{{\Lambda }}}{3}}-\mathscr K(t)\; .
\end{equation}
The initial value of $\mathscr K(t)$ can be obtained by the comparison between \eqref{mfeq1} and \eqref{feq1},
\begin{equation}\label{origin1}
2\mathscr K(t_0)\frac{\dot a(t_0)}{a(t_0)}+\mathscr K(t_0)^2=\frac{{{\Lambda }_{0}}}{3}-\frac{{{\Lambda }}}{3}
\end{equation}
where $t_0\simeq H_0^{-1}$ is the moment at present , or the age of universe now and  $H_0$ is the Hubble constant. Phenomenologically, there are two choices on the initial value of $\mathscr K(t)$, 
\begin{equation}\label{k1}
\mathscr K_1={{H}_{0}}\left(  \sqrt{1-\dfrac{\Lambda -{{\Lambda }_{0}}}{3{{H}_{0}}^{2}}}-1 \right)
\end{equation}
and
\begin{equation}\label{k2}
\mathscr K_2=-{{H}_{0}}\left(  \sqrt{1-\dfrac{\Lambda -{{\Lambda }_{0}}}{3{{H}_{0}}^{2}}}+1 \right)\; .
\end{equation} 
In this scenario, a constrain condition for $\Lambda_0$ must be satisfied as 
\begin{equation}\label{lambda-0}
{\Lambda }_{0}\ge -0.401\Lambda  \approx -\frac{2}{5}\Lambda
\end{equation}
in order to get a reasonable evolution of  $\mathscr K(t)$. 


We can make three kind of approximations listed as Case A, B and C to fix the evolution of $\mathscr K$
on the basis of the second Friedmann equation of $\Lambda$CDM model\eqref{feq2}.

By comparing eq.\eqref{mfeq2} and eq.\eqref{feq2}, we can make the first approximation named {\bf Case A} as
\begin{equation}\label{case1}
\frac{d}{dt}\left( a\mathscr K \right)\text{=}-\frac{1}{3}a\left( \Lambda -{{\Lambda }_{0}} \right)
\end{equation}

Suppose the EoS of cosmic media is ${p(t)=w(t)\rho(t)}$, eq.\eqref{mfeq1} and eq.\eqref{mfeq2} can be converted to the dependence of evolution $a(t)$ on $\mathscr K$ and $\Lambda_0$ 
\begin{equation}\label{ddot a LV}
\frac{\ddot a}{a}+\frac{3w+1}{2}\frac{\dot a^2+k}{a^2}=-\dot{\mathscr K}-\frac{3w+1}{2}\mathscr K^2-(3w+2)\frac{\dot a}{a}\mathscr K+\frac{w+1}{2}{\Lambda }_{0}
\end{equation}
by eliminating $p$ and $\rho$.
Do the same to  eq.\eqref{feq1} and eq.\eqref{feq2}, we get the dependence of evolution $a(t)$ on $\Lambda$ as
\begin{equation}\label{ddot a CDM}
\frac{\ddot a}{a}+\frac{3w+1}{2}\frac{\dot a^2+k}{a^2}=\frac{w+1}{2}\Lambda
\end{equation}

We can make the second approximation named {\bf Case B} by requiring 
\begin{equation}\label{case2}
\dot{\mathscr K}+\frac{3w+1}{2}\mathscr K^2+(3w+2)\frac{\dot a}{a}\mathscr K=\frac{w+1}{2}(\Lambda_0-\Lambda)\; ,
\end{equation}
which relates contortion to $w(t)$. 

Suppose the dark partner satisfies equation of state ${{p}_{K}}\text{=}{{w}_{0}}{{\rho }_{K}}$, we make the third approximation named {\bf Case C} as
\begin{equation}\label{case3}
\left( 3{{w}_{0}}+1 \right){{\mathscr K}^{2}}+\left( 6{{w}_{0}}+4 \right)\frac{{\dot{a}}}{a}\mathscr K+2\dot{\mathscr K}=\left( {{w}_{0}}+1 \right){{\Lambda }_{0}}\;.
\end{equation}
from eq.\eqref{rhoK} and eq.\eqref{pK}.

The evolution of $H(t)$ and $\mathscr K(t)$ can be determined by the equations \eqref{ddot a LV} together with one of the approximations eq.\eqref{case1}, eq.\eqref{case2} and eq.\eqref{case3} and with the initial conditions $H(t_0)=H_0$ and $\mathscr K(t_0)=\mathscr K_1$ of \eqref{k1} or $\mathscr K(t_0)=\mathscr K_2$ of \eqref{k2}.

\begin{table}[!htbp]
	\centering
	\caption{Models of Approximation in Large Scale Lorentz Violation Cosmology}\label{model}
	\begin{tabular}{ccc}
		\toprule
		&Values of $\mathscr K(t_0)$
		&Evolution Eq. of $\mathscr K(t)$\\
		\midrule
		{\bf Case A1}
		&$\mathscr K(t_0)=\mathscr K_1$
		&\multirow{2}{*}{$\dfrac{d}{dt}\left( a\mathscr K \right)\text{=}-\dfrac{1}{3}a\left( \Lambda -{{\Lambda }_{0}} \right)$}\\
		{\bf Case A2}
		&$\mathscr K(t_0)=\mathscr K_2$
		& \\
		\hline
		{\bf Case B1}
		&$\mathscr K(t_0)=\mathscr K_1$
		&\quad\multirow{2}{*}{$\dot{\mathscr K}+\dfrac{3w+1}{2}\mathscr K^2+(3w+2)\dfrac{\dot a}{a}\mathscr K=\dfrac{w+1}{2}(\Lambda_0-\Lambda)$}\\
		{\bf Case B2}
		&$\mathscr K(t_0)=\mathscr K_2$
		& \\
		\hline
		{\bf Case C1}
		&$\mathscr K(t_0)=\mathscr K_1$
		&\multirow{2}{*}{$\left( 3{{w}_{0}}+1 \right){{\mathscr K}^{2}}+\left( 6{{w}_{0}}+4 \right)\dfrac{{\dot{a}}}{a}\mathscr K+2\dot{\mathscr K}=\left( {{w}_{0}}+1 \right){{\Lambda }_{0}}$}\\
		{\bf Case C2}
		&$\mathscr K(t_0)=\mathscr K_2$
		& \\
		\bottomrule
	\end{tabular}
\end{table}

Table \ref{model} summarizes the models of approximation discussed above.


\begin{figure}[!htbp] 
	\centering	
	\subfigure[]
	{\label{htk0pw0_m089m1l0_m02l}
		\includegraphics[width=3in]{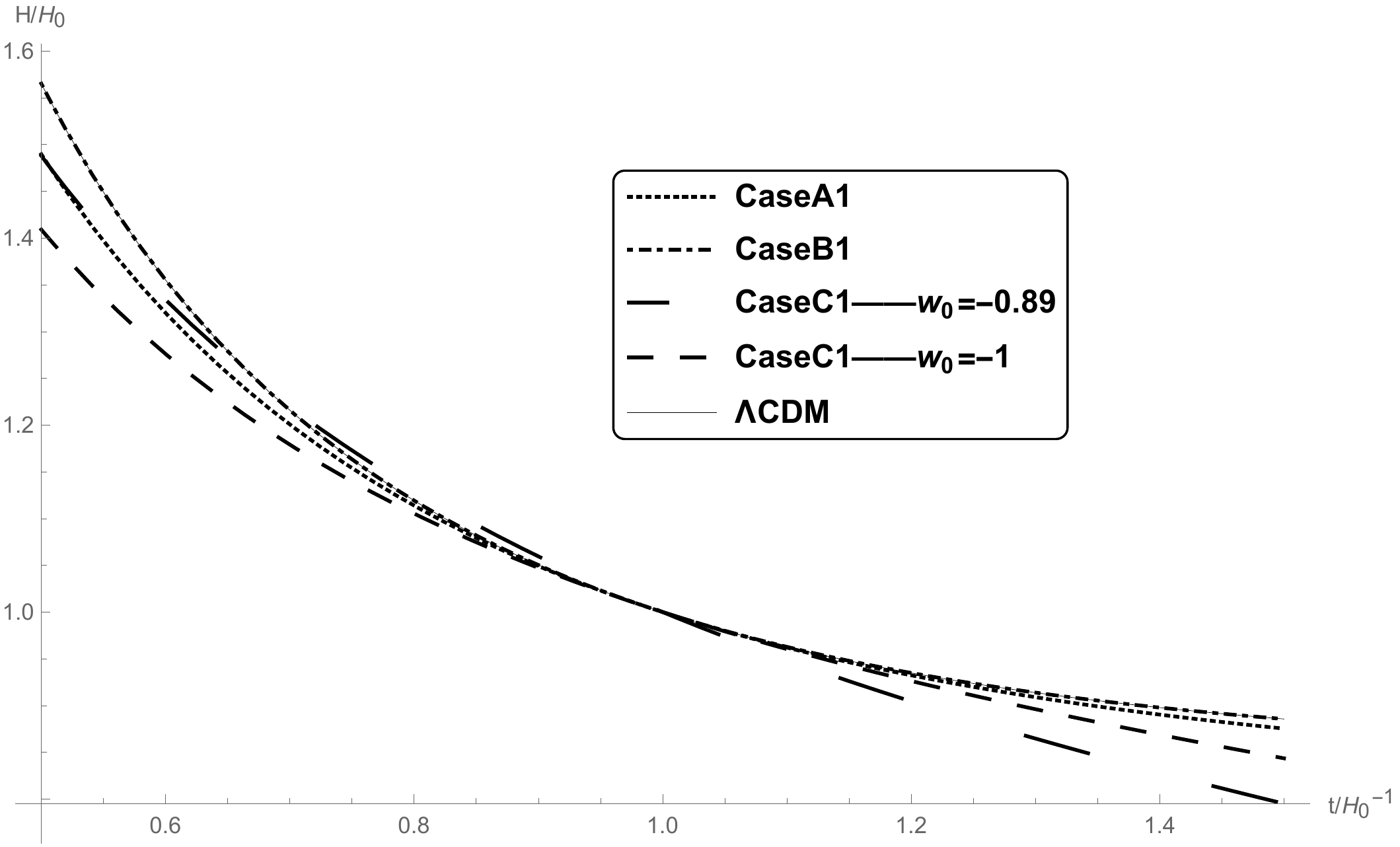}
	}
	\subfigure[]
	{\label{htk0pw0_m089m1l0_0l}
		\includegraphics[width=3in]{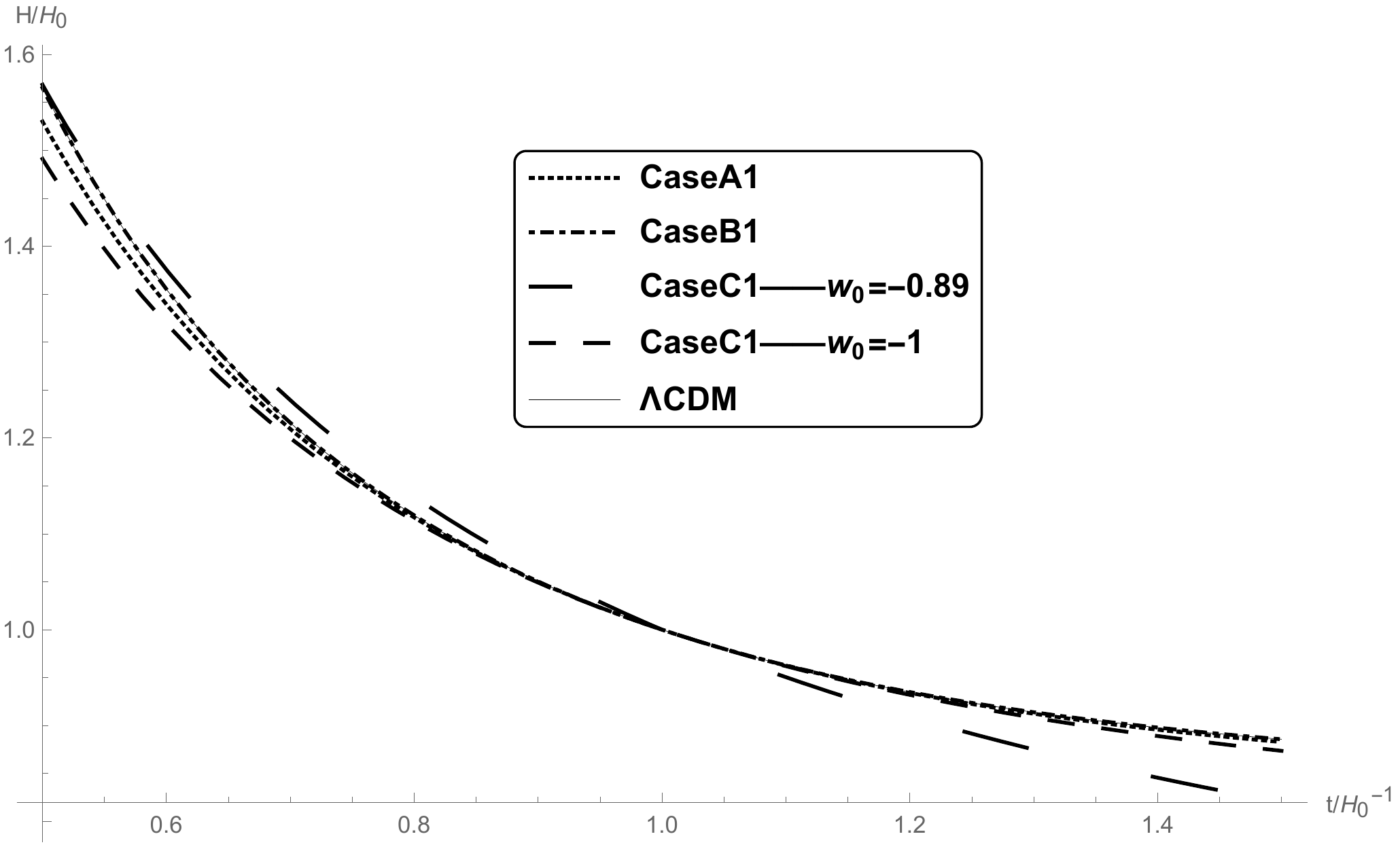}
	}
	\subfigure[]
	{\label{htk0pw0_m089m1l0_02l}
		\includegraphics[width=3in]{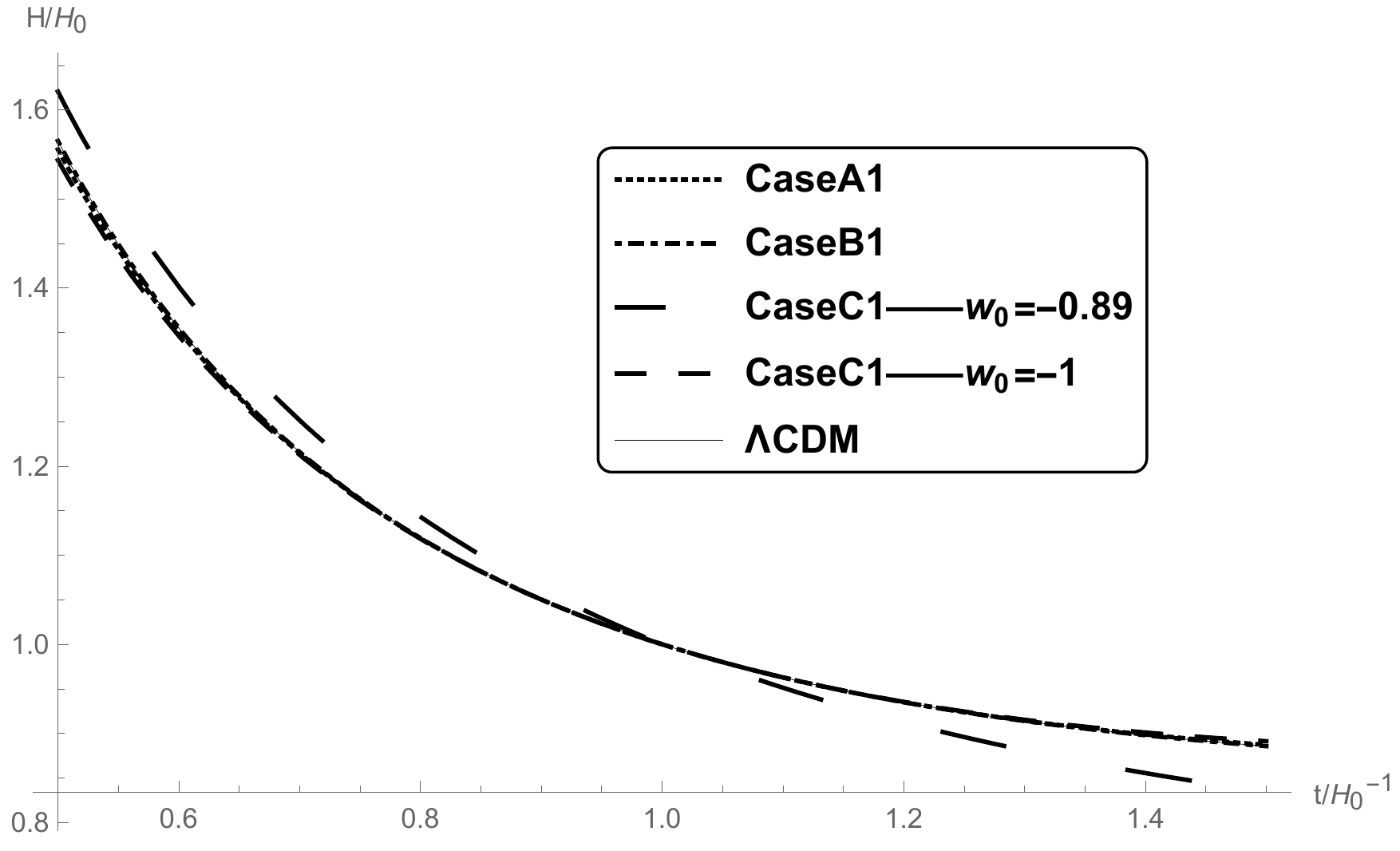}
	}
	\subfigure[]
	{\label{htk0mw0_m089m1l0_m02l}
		\includegraphics[width=3in]{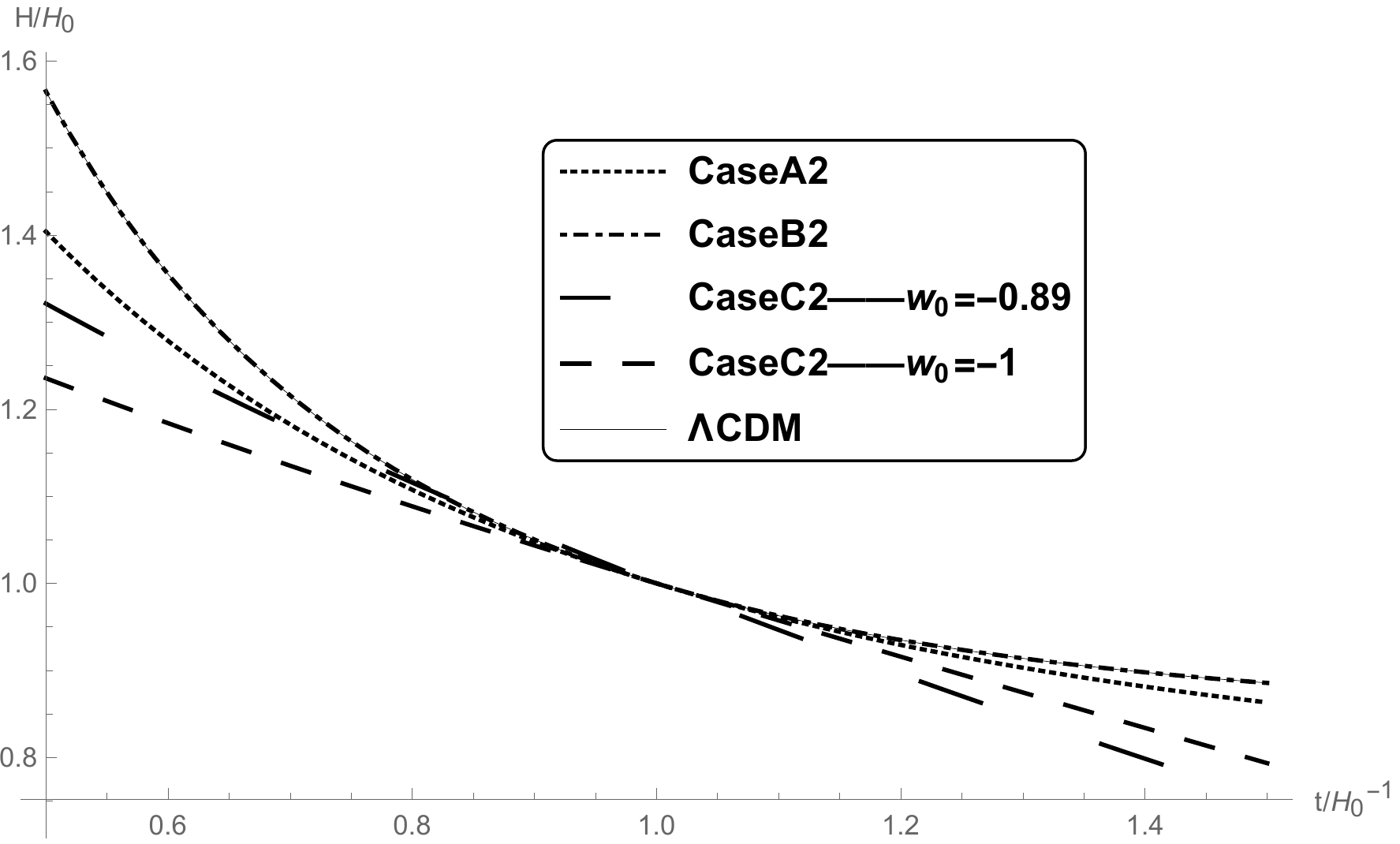}
	}
	\subfigure[]
	{\label{htk0mw0_m089m1l0_0l}
		\includegraphics[width=3in]{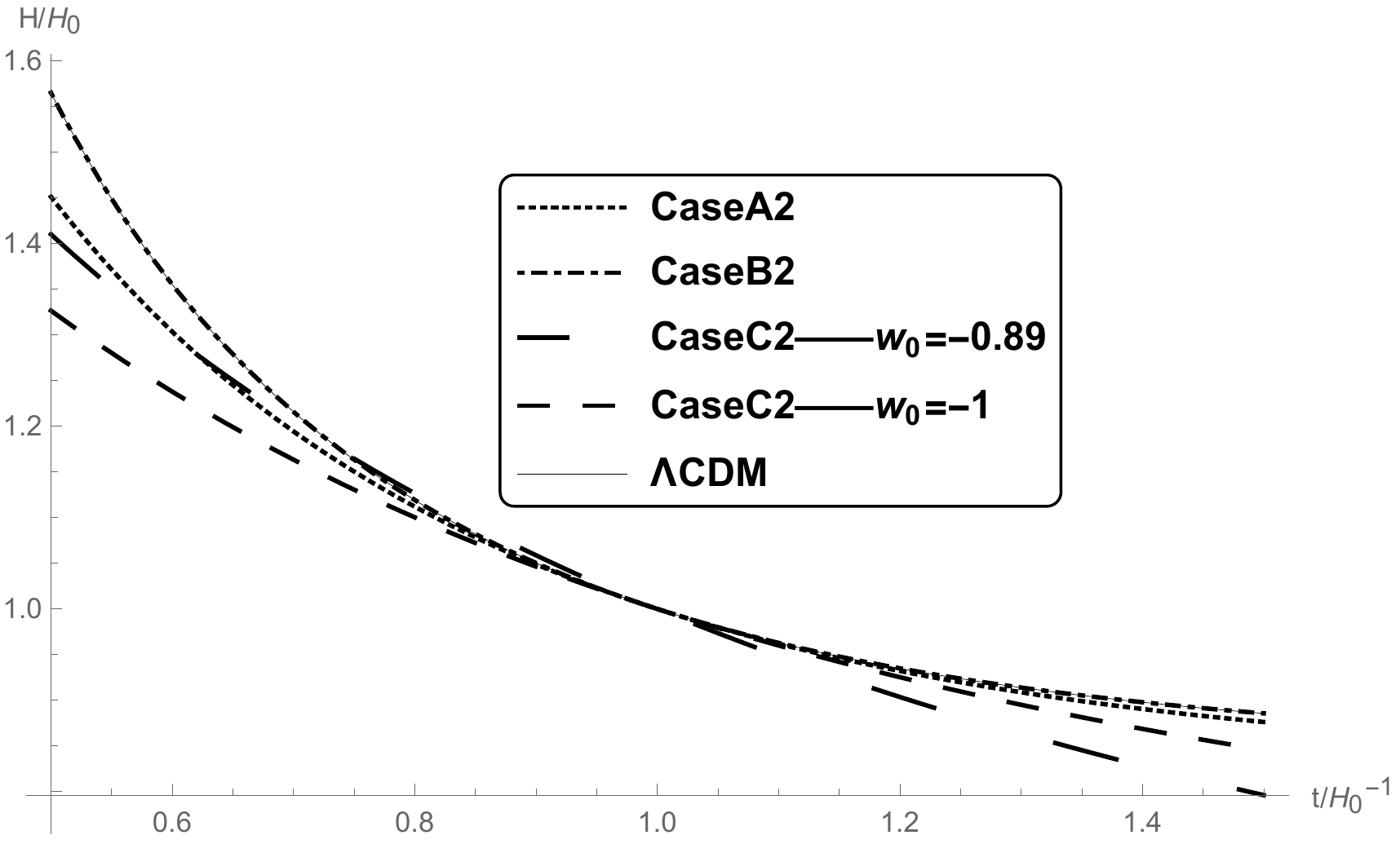}
	}
	\subfigure[]
	{\label{htk0mw0_m089m1l0_02l}
		\includegraphics[width=3in]{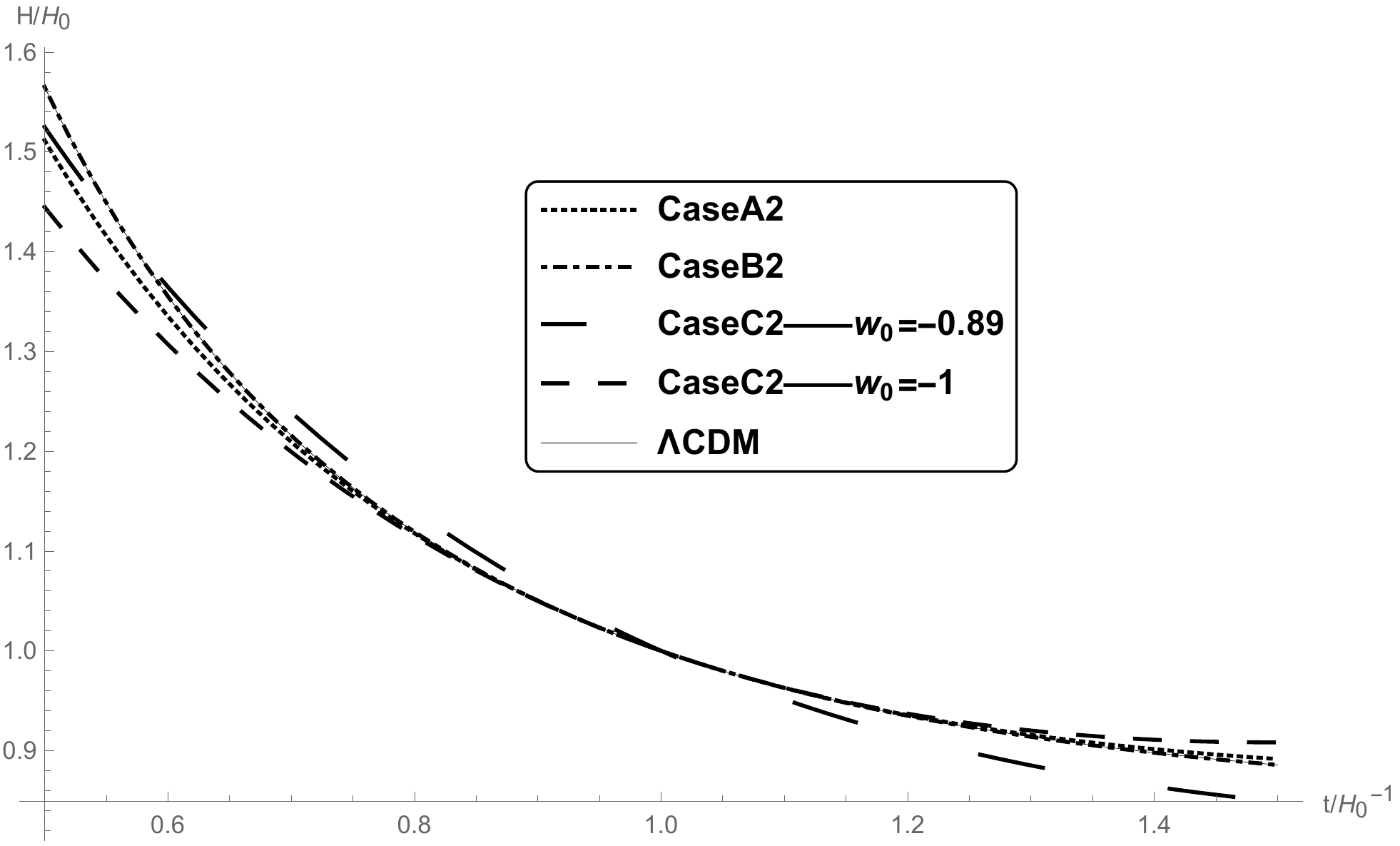}
	}
	\caption{The evolution of Hubble constant $H$ versus $t$ in the case of initial value $\mathscr K\left( {{t}_{0}} \right)={{H}_{0}}\left( \pm \sqrt{1-\frac{\Lambda -{{\Lambda }_{0}}}{3{{H}_{0}}^{2}}}-1 \right)$ and among three case of approximation and $\Lambda CDM$ model, (a) and (d) $\Lambda_0=-0.2\Lambda$, (b) and (e) $\Lambda_0=0$ , (c) and (f)  $\Lambda_0=0.2\Lambda$ }\label{htk0pmw0_m089m1}
\end{figure}

\begin{figure}[!htbp] 
	\centering	
	\subfigure[]
	{\label{ktk0pw0_m089m1l0_m02l}
		\includegraphics[width=3in]{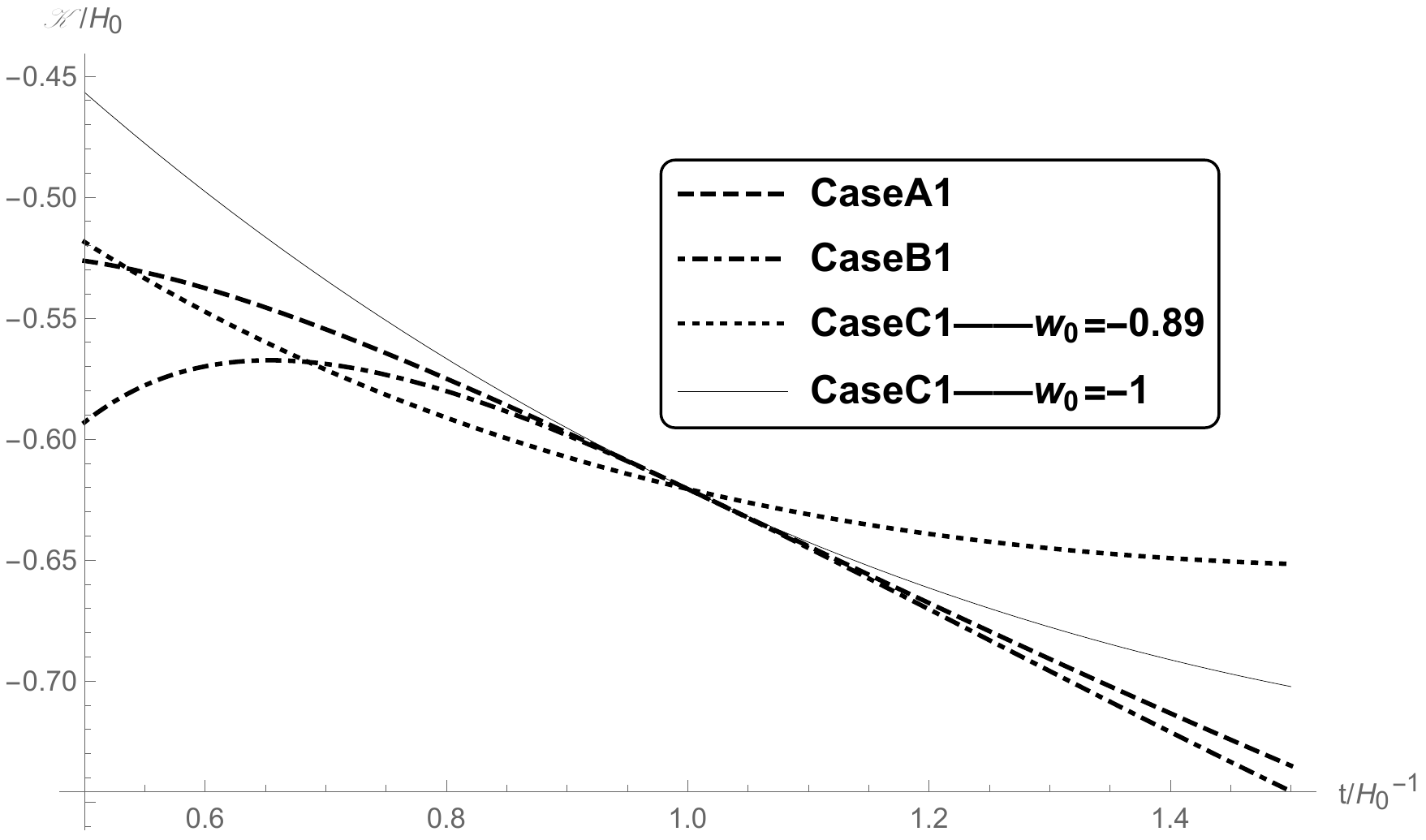}
	}
	\subfigure[]
	{\label{ktk0pw0_m089m1l0_0l}
		\includegraphics[width=3in]{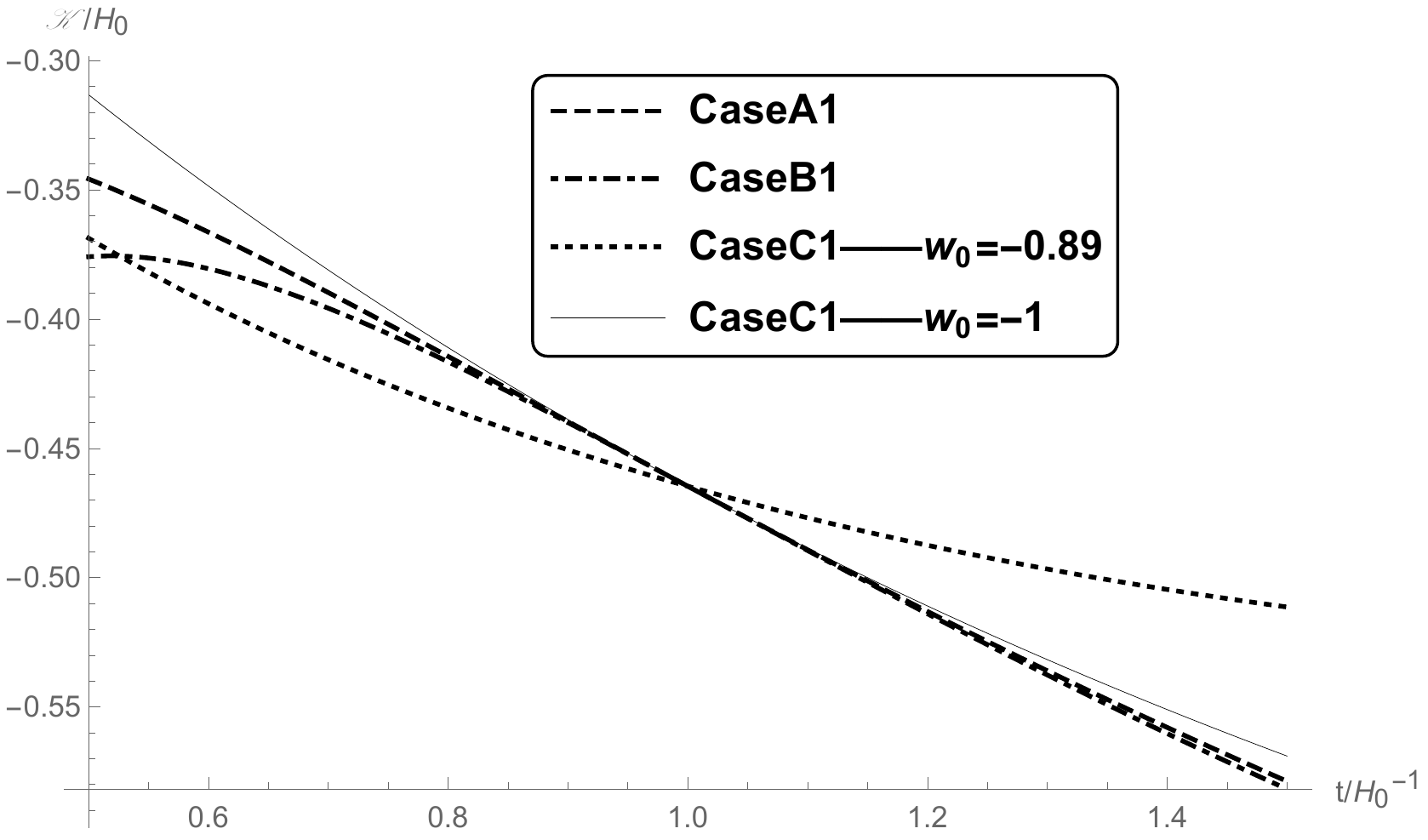}
	}
	\subfigure[]
	{\label{ktk0pw0_m089m1l0_02l}
		\includegraphics[width=3in]{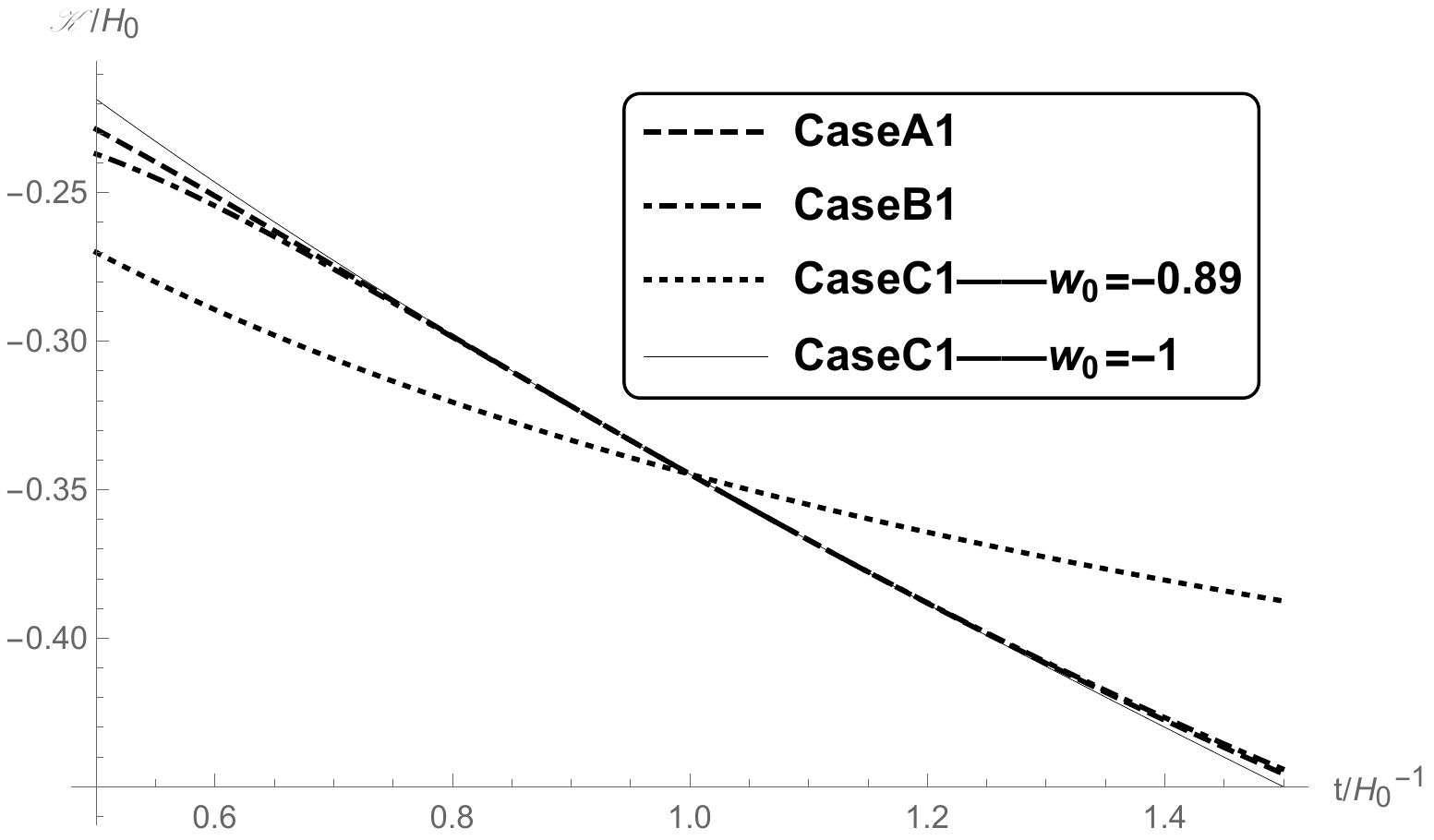}
	}
	\subfigure[]
	{\label{ktk0mw0_m089m1l0_m02l}
		\includegraphics[width=3in]{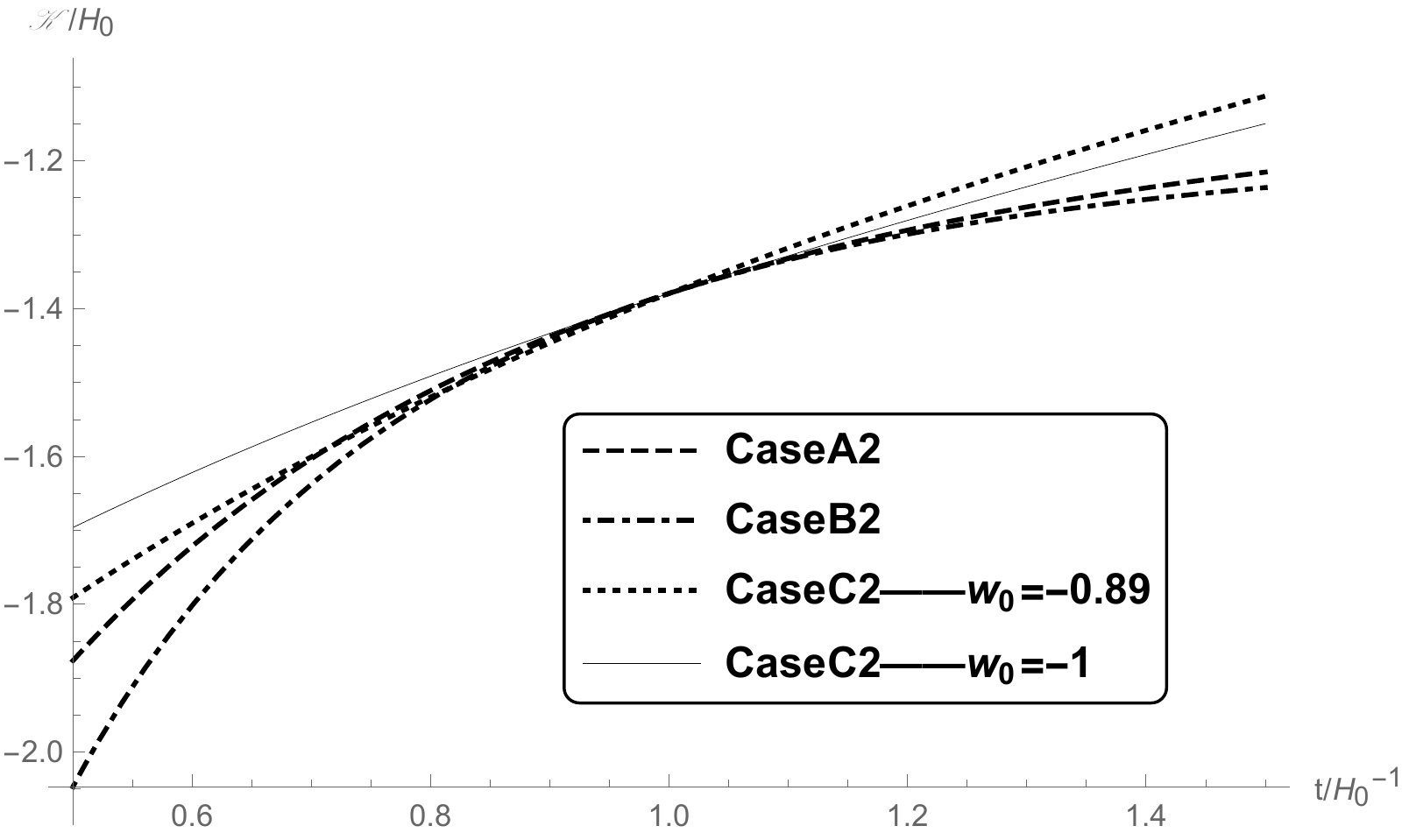}
	}
	\subfigure[]
	{\label{ktk0mw0_m089m1l0_0l}
		\includegraphics[width=3in]{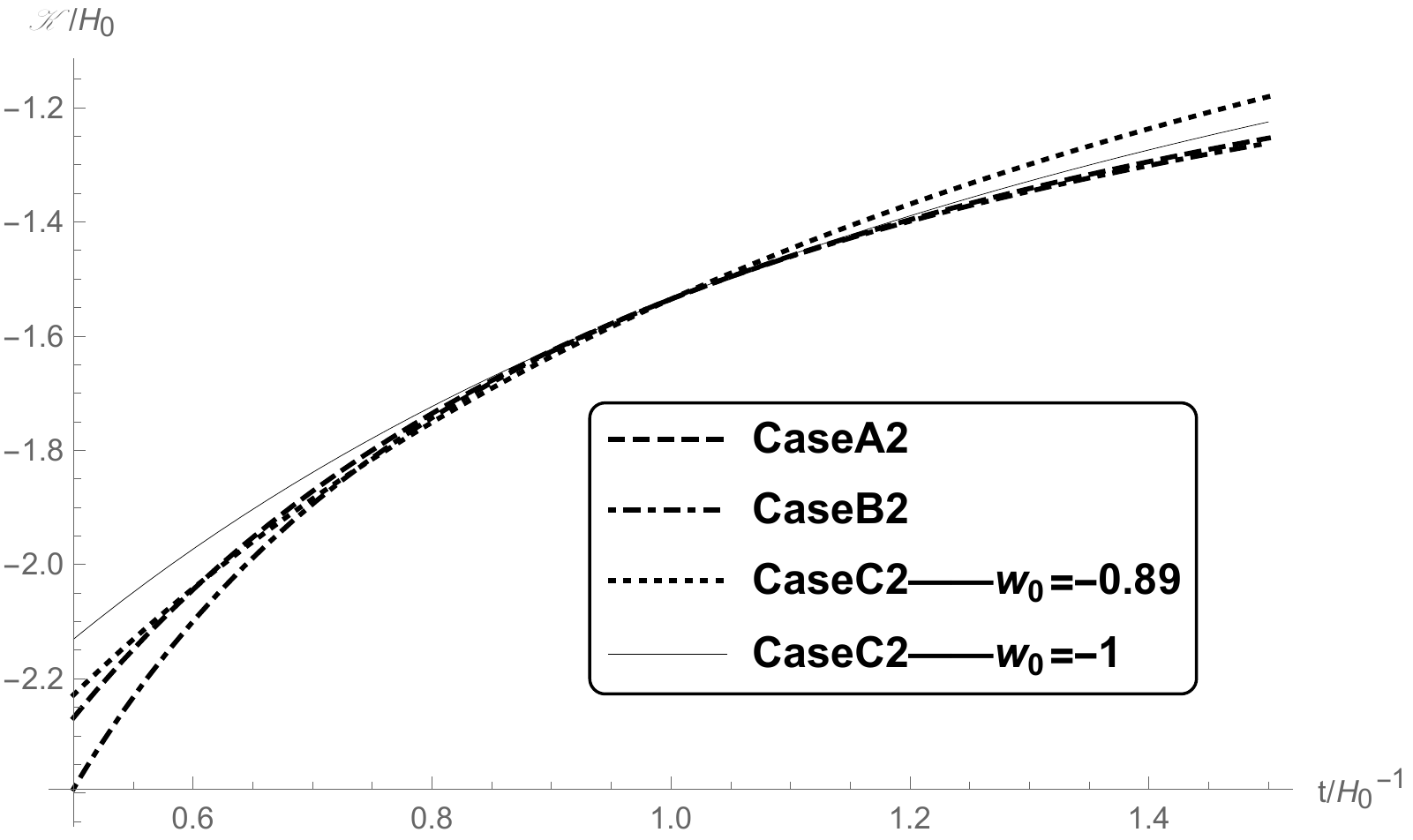}
	}
	\subfigure[]
	{\label{ktk0mw0_m089m1l0_02l}
		\includegraphics[width=3in]{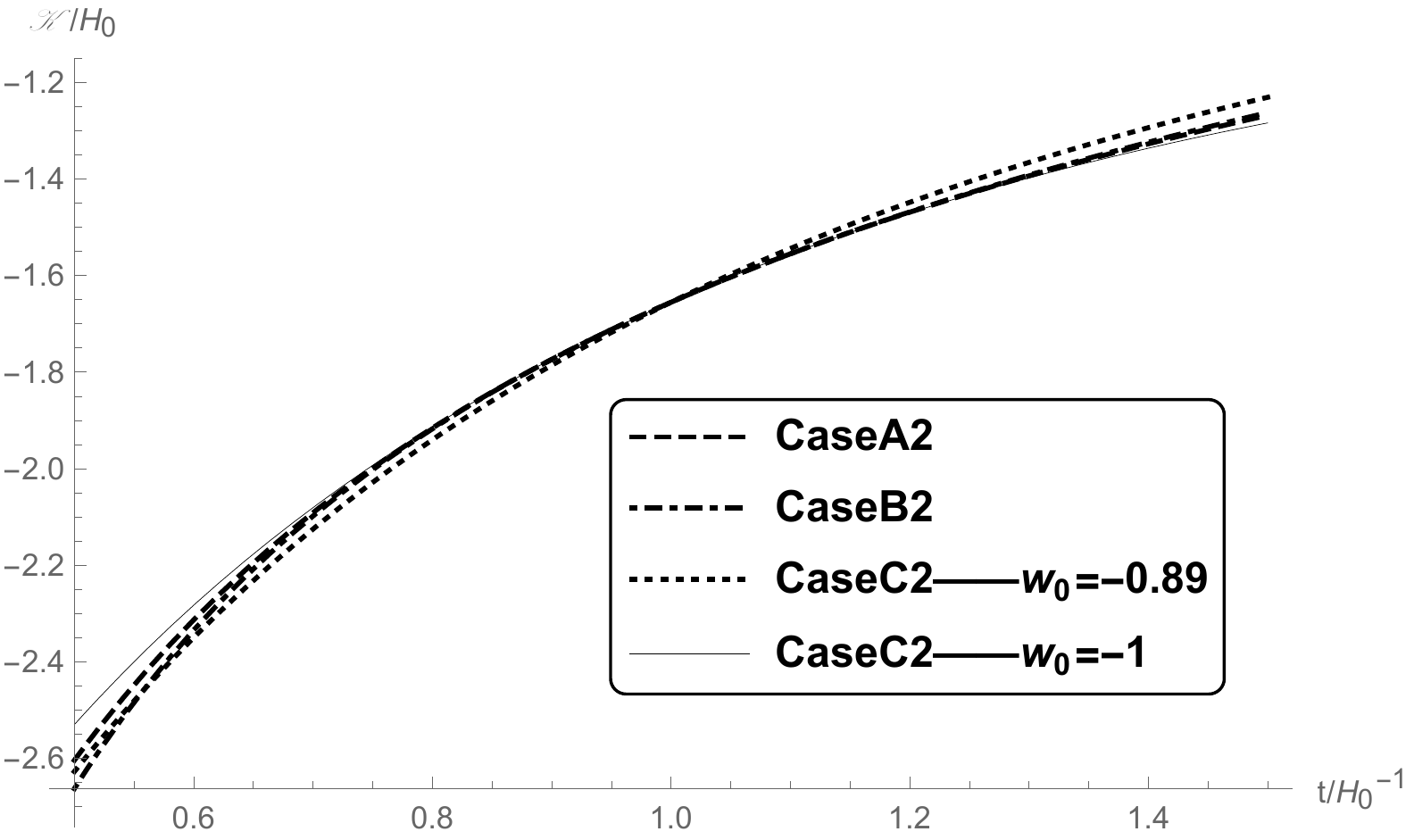}
	}
	\caption{The evolution of $\mathscr K$ versus $t$ in the case of initial value $\mathscr K\left( {{t}_{0}} \right)={{H}_{0}}\left( \pm \sqrt{1-\frac{\Lambda -{{\Lambda }_{0}}}{3{{H}_{0}}^{2}}}-1 \right)$ among three case of approximation, (a) and (d) $\Lambda_0=-0.2\Lambda$, (b) and (e) $\Lambda_0=0$ , (c) and (f)  $\Lambda_0=0.2\Lambda$ }\label{ktk0pmw0_m089m1}
\end{figure}

The evolution of $H(t)$ and $\mathscr K(t)$ versus $t$ and  ${{\Lambda }_{0}}$ in all the cases of approximations are presented in Fig.\ref{htk0pmw0_m089m1} and Fig.\ref{ktk0pmw0_m089m1} where  we set ${w(t)\simeq 0}$ in the EoS of the cosmic media for cold matter dominates the energy density in the period of late time expansion of the universe and $w_0=-0.89$ and $w_0=-1$ in Case C as examples.

It is apparent that the evolutions of Hubble constant versus $t$ in these models are very close and ones of {\bf Case B} and $\Lambda CDM$ model tend to almost coincide with each other in the long time evolution. There is clear but small difference between the longtime evolution curves of {\bf Case A} and $\Lambda CDM$ model for $\Lambda_0 < 0$ while one between {\bf Case C} at $w_0=-1$ and $\Lambda CDM$ model is a little bit larger in the initial value $\mathscr K\left( {{t}_{0}} \right)={{H}_{0}}\left( \sqrt{1-\dfrac{\Lambda -{{\Lambda }_{0}}}{3{{H}_{0}}^{2}}}-1 \right)$  case. The differences among the evolution curve for $H(t)$ decrease with the increase of $\Lambda_0$ so that ones for {\bf Case A}, {\bf Case B}, {\bf Case C} and $\Lambda CDM$ model are almost the same. In the long time evolution of Hubble constant, the difference between {\bf Case A} and $\Lambda CDM$ model decreases as the increase of $\Lambda_0$ while one between {\bf Case C} at $w_0=-\dfrac{2}{3}$ and $\Lambda CDM$ model is getting larger more and more. The case for $w_0=-\dfrac{1}{3}$ of {\bf Case C} is similar to case of $w_0=-\dfrac{2}{3}$ while only differes at the deviation from $\Lambda CDM$ model is larger than the $w_0=-\dfrac{2}{3}$ case.

In the initial value $\mathscr K\left( {{t}_{0}} \right)=-{{H}_{0}}\left( \sqrt{1-\frac{\Lambda -{{\Lambda }_{0}}}{3{{H}_{0}}^{2}}}\text{+}1 \right)$ case, the difference between the time evolution of {\bf Case A} and $\Lambda CDM$ model is always smaller than one between {\bf Case C} and $\Lambda CDM$ model no matter what $\Lambda_0$ is. The evolution curves of these several models coincide almost identically where $\Lambda_0$ value is around zero. The deviations of both {\bf Case A} and {\bf Case C} from $\Lambda CDM$ model increase along with the increase of $\Lambda_0$ value while one for {\bf Case C} is getting larger than {\bf Case A}.

The evolution of $\mathscr K(t)$ versus $t$ and  ${{\Lambda }_{0}}$  are presented in Fig.\ref{ktk0pmw0_m089m1} where the results of {\bf Case A} and {\bf Case B} are very close but one of {\bf Case C} deviates a little bit larger from one of {\bf Case A} and {\bf Case B}. Especially in the initial value $\mathscr K\left( {{t}_{0}} \right)={{H}_{0}}\left( \sqrt{1-\dfrac{\Lambda -{{\Lambda }_{0}}}{3{{H}_{0}}^{2}}}-1 \right)$ case, results of {\bf Case A} and {\bf Case B} are almost the same while one of {\bf Case C} differs. We can find that {\bf Case C} is not a good approximation probably for the reason that $w_0$ need to evolve either.

\begin{figure}[!htbp]

	\centering	
	\subfigure[]
	{\label{dlk0pw0_m089m1l0_m02l}
		\includegraphics[width=3in]{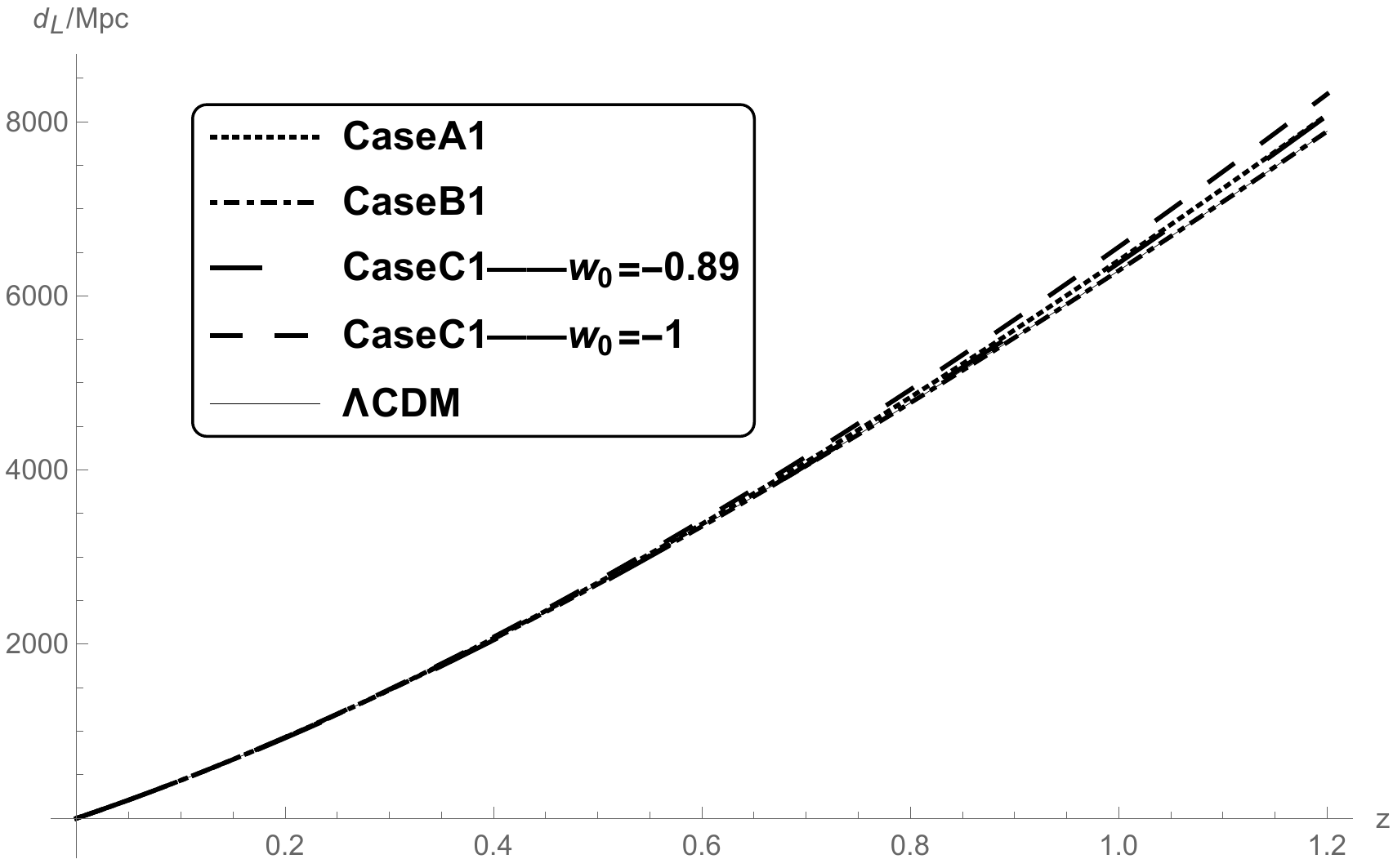}
	}
	\subfigure[]
	{\label{dlk0pw0_m089m1l0_0l}
		\includegraphics[width=3in]{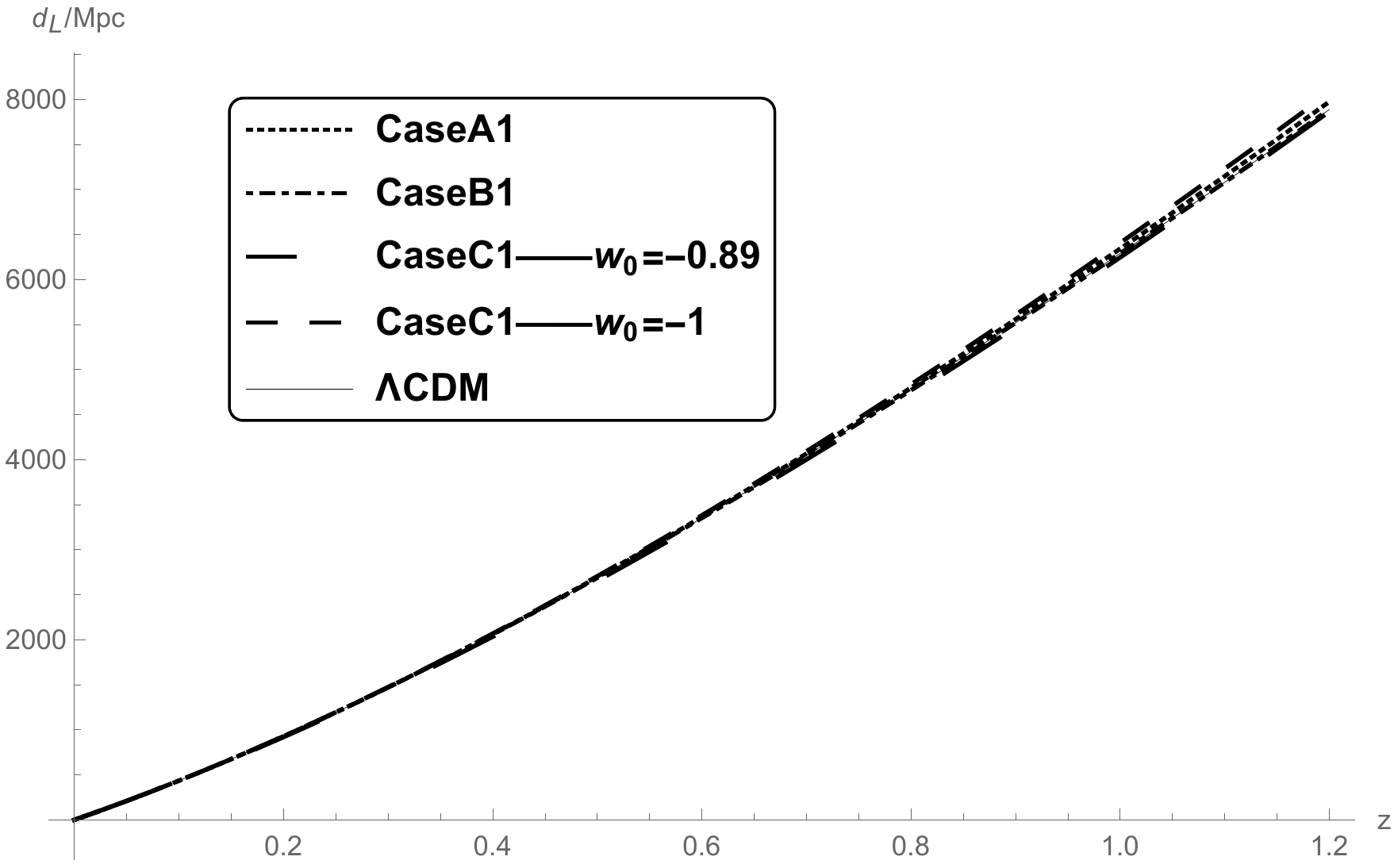}
	}
	\subfigure[]
	{\label{dlk0pw0_m089m1l0_02l}
		\includegraphics[width=3in]{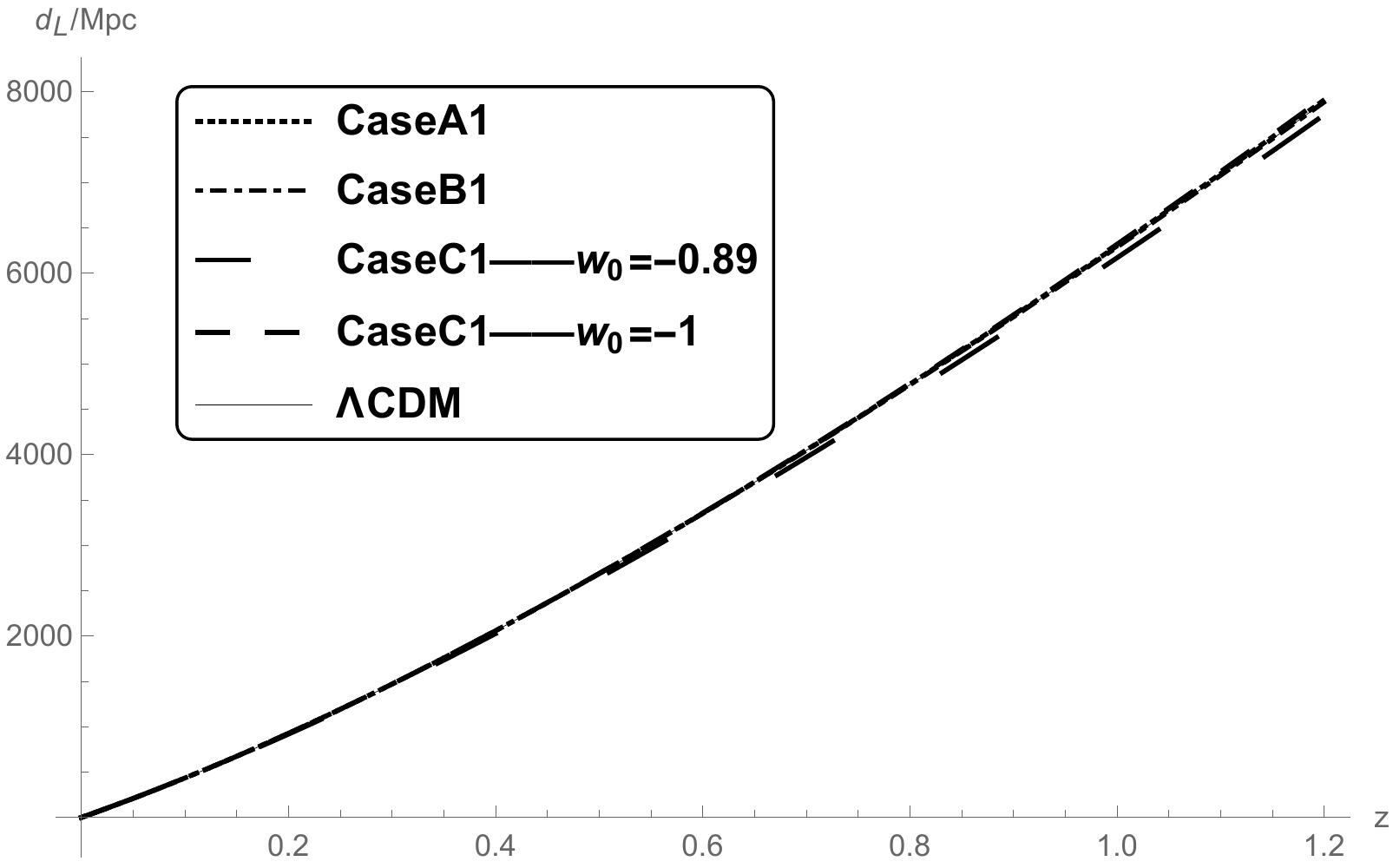}
	}
	\subfigure[]
	{\label{dlk0mw0_m089m1l0_m02l}
		\includegraphics[width=3in]{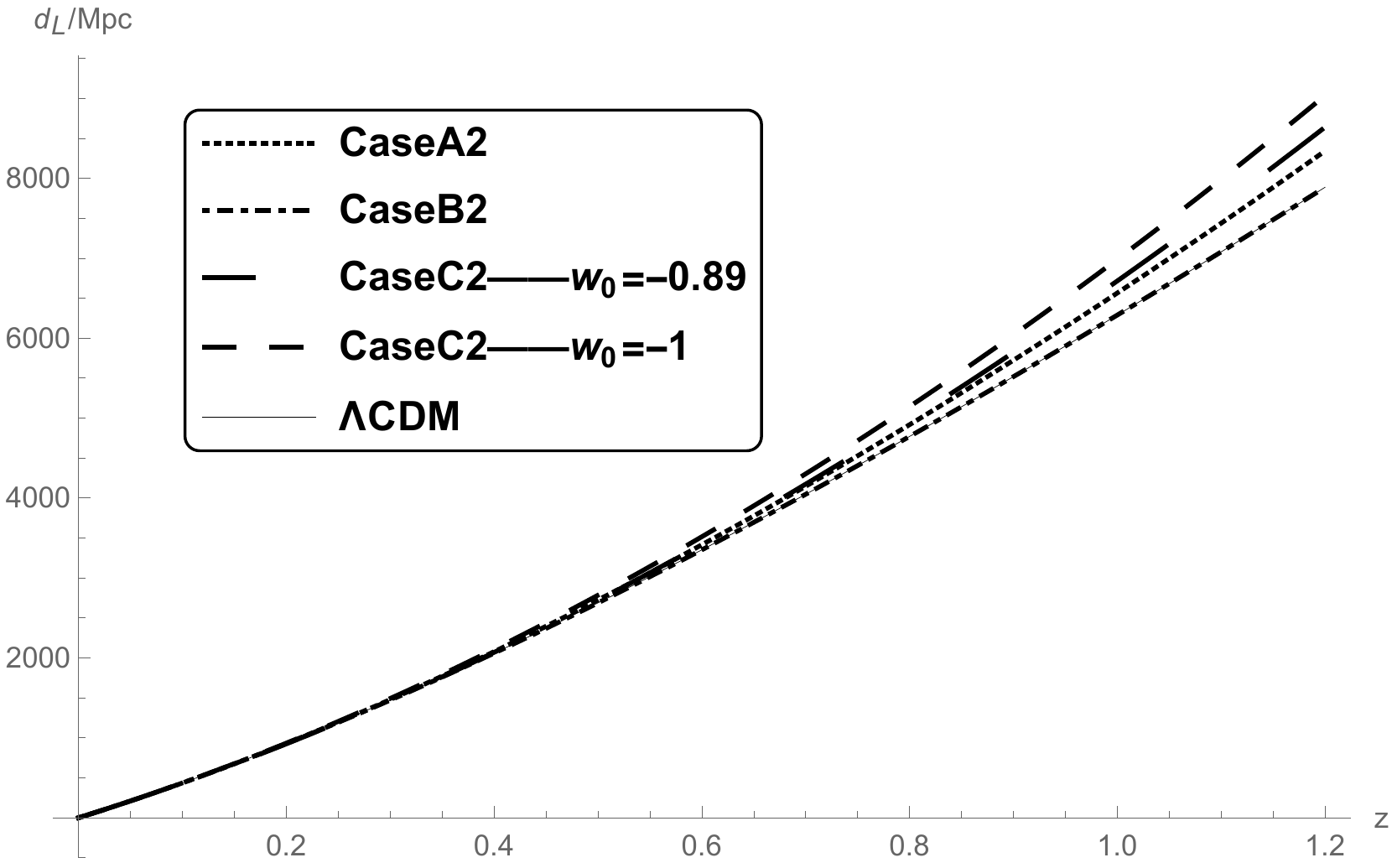}
	}
	\subfigure[]
	{\label{dlk0mw0_m089m1l0_0l}
		\includegraphics[width=3in]{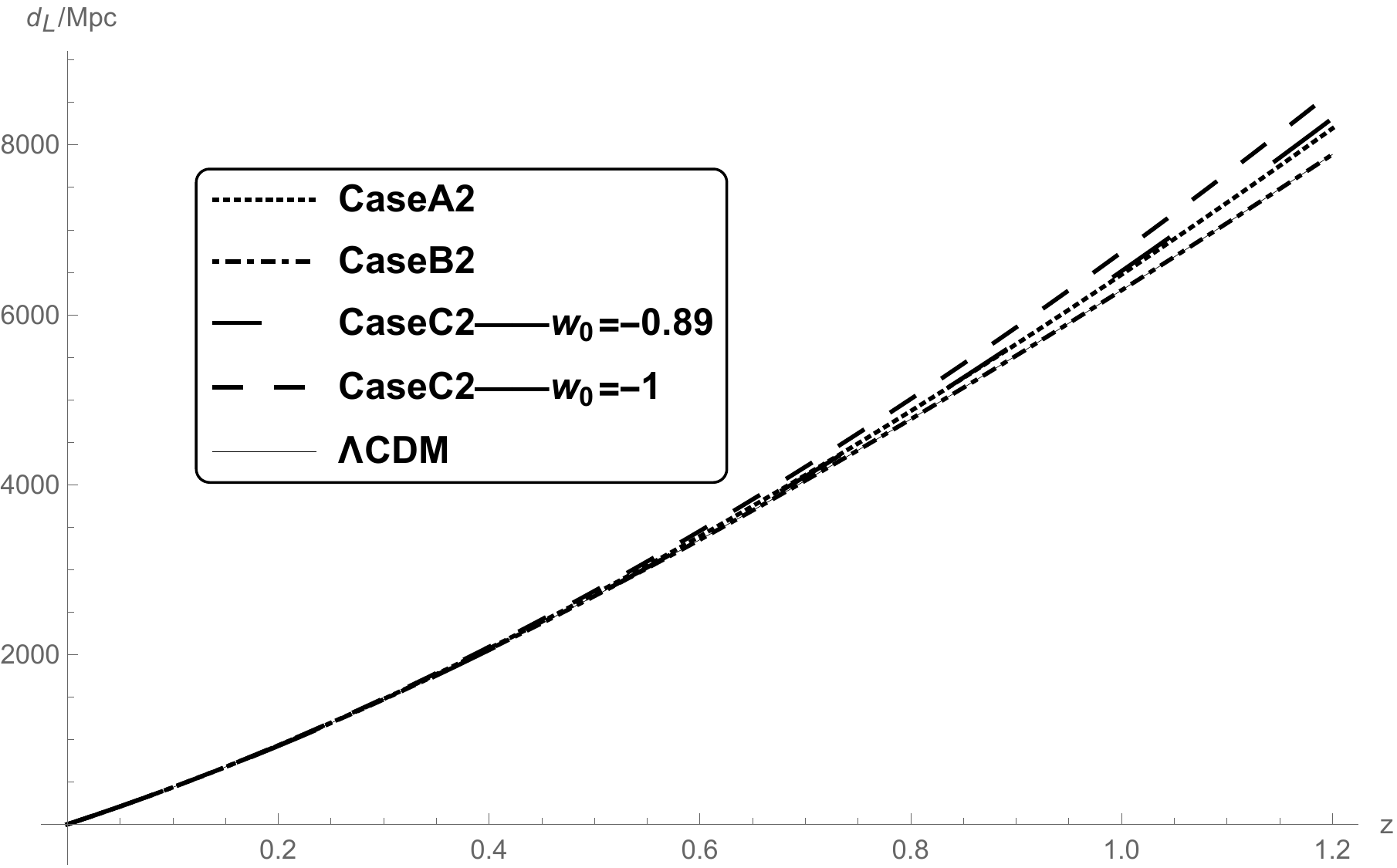}
	}
	\subfigure[]
	{\label{dlk0mw0_m089m1l0_02l}
		\includegraphics[width=3in]{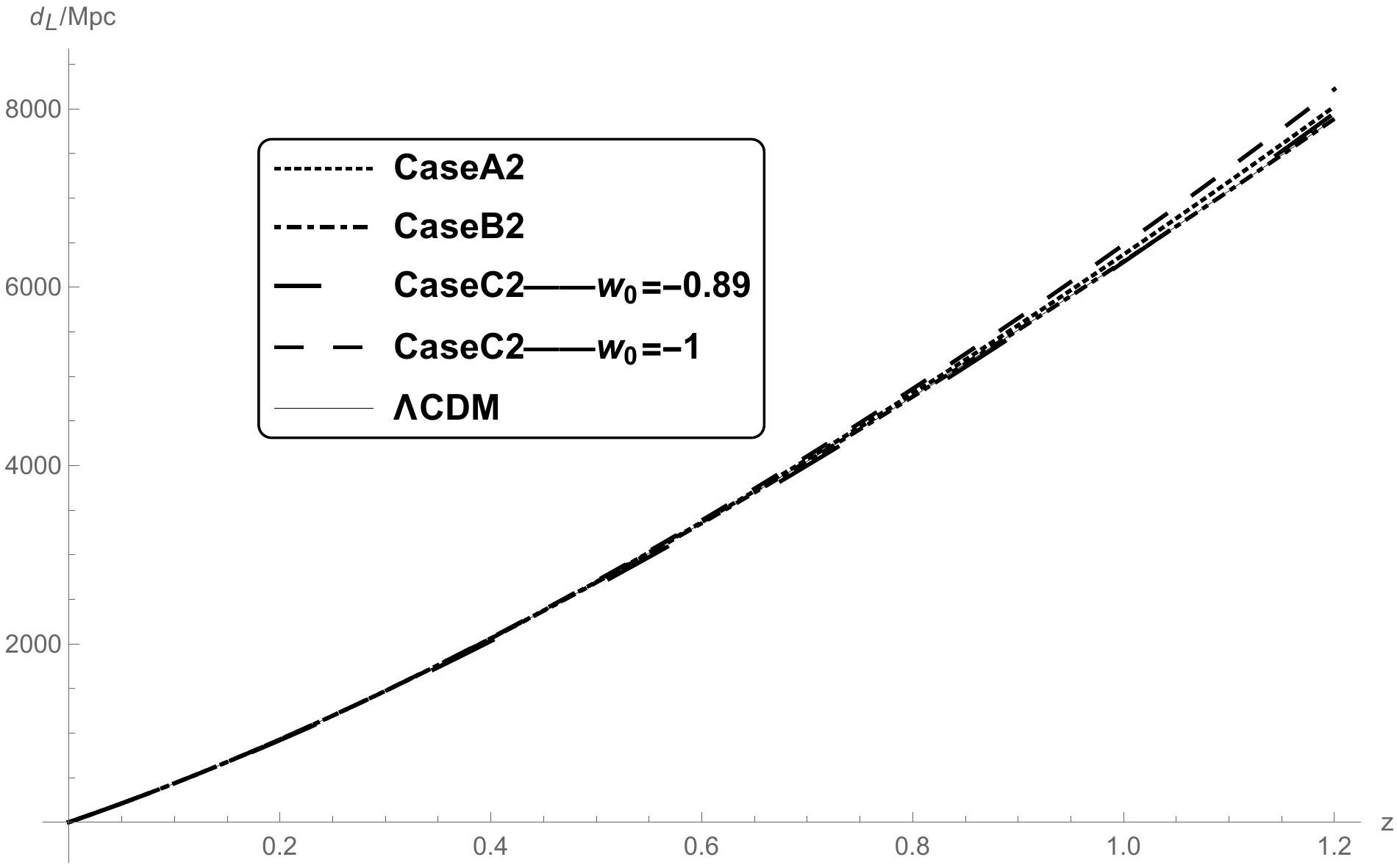}
	}
	\caption{the comparison of luminosity distance in the case of initial value $\mathscr K\left( {{t}_{0}} \right)={{H}_{0}}\left( \pm \sqrt{1-\frac{\Lambda -{{\Lambda }_{0}}}{3{{H}_{0}}^{2}}}-1 \right)$ among three case of approximation and $\Lambda CDM$ model, (a) and (d) $\Lambda_0=-0.2\Lambda$, (b) and (e) $\Lambda_0=0$ , (c) and (f)  $\Lambda_0=0.2\Lambda$ }\label{dlk0pmw0_m089m1}
\end{figure}

In the case of non-zero contortion, the world line of a photon is still the light-like geodesic curve as in the Riemann space-time case rather than the auto-parallel curve. The redshift formula is the same as in the Lorentz invariant zero contortion case, 
\begin{equation}\label{za}
1+z=\frac{{{a}_{0}}}{a}
\end{equation}
and 
\begin{equation}\label{dzda}
\frac{dz}{da}=-\frac{{{a}_{0}}}{{{a}^{2}}}\;  
\end{equation}
where $a_0=a(t_0)$. We can convert $H$ and $\mathscr K$ versus $t$ to $z$,
\begin{equation}\label{dhdz}
\dot{H}\left( t \right)=H'\left( z \right)\left( -\frac{{{a}_{0}}}{{{a}^{2}}} \right)aH=-\left( 1+z \right)H\left( z \right)H'\left( z \right)
\end{equation}
and
\begin{equation}\label{dkdz}
\dot{\mathscr K}\left( t \right)=\frac{d\mathscr K}{da}\frac{da}{dt}=\mathscr K'\left( a \right)aH=-\left( 1+z \right)H\left( z \right)\mathscr K'\left( z \right)
\end{equation}
with relations
\begin{equation}\label{dhdt}
\dot{H}\left( t \right)=\frac{dH}{dt}=\frac{dH}{da}\frac{da}{dt}=H'\left( a \right)aH
\end{equation}
and 
\begin{equation}\label{dhda}
H'\left( a \right)=\frac{dH}{da}=\frac{dH}{dz}\frac{dz}{da}=H'\left( z \right)\frac{dz}{da}
\end{equation}
By definition of luminosity distance $d_L$ (see \cite{Weinberg:2008zzc}) we can get
\begin{equation}\label{hzdl}
d_L(z)=\left(1+z\right)\int^z_0\frac{1}{H(z')}\,\mathrm dz'\quad
\end{equation}
in the case of $k=0$ and 
\begin{equation}\label{dtdz}
\frac{\mathrm dt}{\mathrm dz}=-\frac{1}{1+z}\,\frac{\mathrm d}{\mathrm dz}\left(\frac{d_L}{1+z}\right)
\end{equation}
With eq.\eqref{hzdl} and eq.\eqref{dtdz}, we can convert Friedmann equations \eqref{ddot a LV} to equation for $d_L(z)$ and $\mathscr K(z)$ versus redshift $z$,
\begin{multline}\label{dlkzmfeq}
\frac{{{\left( 1+z \right)}^{6}}{{d}_{L}}^{\prime \prime }\left( z \right)}{{{\left( \left( 1+z \right){{d}_{L}}^{\prime }\left( z \right)-{{d}_{L}}\left( z \right) \right)}^{3}}}-\frac{1}{2}\frac{{{\left( 1+z \right)}^{4}}}{{{\left( \left( 1+z \right){{d}_{L}}^{\prime }\left( z \right)-{{d}_{L}}\left( z \right) \right)}^{2}}}-\frac{{{\left( 1+z \right)}^{3}}{\mathscr K}'\left( z \right)}{\left( 1+z \right){{d}_{L}}^{\prime }\left( z \right)-{{d}_{L}}\left( z \right)} \\ 
+\frac{\text{2}{{\left( 1+z \right)}^{2}}}{\left( 1+z \right){{d}_{L}}'\left( z \right)-{{d}_{L}}\left( z \right)}\mathscr K\left( z \right)+\frac{1}{2}{{\mathscr K}^{\text{2}}}\left( z \right)\text{=}{{\Lambda }_{0}} \\ 
\end{multline}
So can we do to three approximations eq.\eqref{case1}, eq.\eqref{case2}  and eq.\eqref{case3} to get
\begin{equation}\label{dlkzcase1}
\frac{{{\left( 1+z \right)}^{2}}}{\left( 1+z \right){{d}_{L}}'\left( z \right)-{{d}_{L}}\left( z \right)}\mathscr K\left( z \right)-\frac{{{\left( 1+z \right)}^{3}}}{\left( 1+z \right){{d}_{L}}^{\prime }\left( z \right)-{{d}_{L}}\left( z \right)}{\mathscr K}'\left( z \right)=\frac{1}{3}\left( {{\Lambda }_{0}}-\Lambda  \right)\; ,
\end{equation} 
\begin{equation}\label{dlkzcase2}
\frac{4{{\left( 1+z \right)}^{2}}\mathscr K\left( z \right)-2{{\left( 1+z \right)}^{3}}{K}'\left( z \right)}{\left( 1+z \right){{d}_{L}}'\left( z \right)-{{d}_{L}}\left( z \right)}+{{\mathscr K}^{2}}\left( z \right)\text{=}{{\Lambda }_{0}}-\Lambda 
\end{equation}
and
\begin{equation}\label{dlkzcase3}
\frac{2{{\left( 1+z \right)}^{2}}\left( \left( 3{{w}_{0}}+2 \right)\mathscr K\left( z \right)-(1+z){\mathscr K}'\left( z \right) \right)}{\left( 1+z \right){{d}_{L}}'\left( z \right)-{{d}_{L}}\left( z \right)}+\left( 3{{w}_{0}}+1 \right)\mathscr K{{\left( z \right)}^{2}}=\left( {{w}_{0}}+1 \right){{\Lambda }_{0}}
\end{equation}
Comparisons of the luminosity distance $d_L$ curve versus redshift $z$ among three models of approximation and $\Lambda$CDM model are presented in Fig.\ref{dlk0pmw0_m089m1}. 

It is observed that no matter what value ${{w}_{0}}$ and ${{\Lambda }_{0}}$ take, the luminosity distance $d_L$ versus redshift $z$ curves of {\bf Case B} and $\Lambda CDM$ model appear to coincide with each other almost. The difference between {\bf Case A} and $\Lambda CDM$ model and one between {\bf Case C} at $w_0=-1$ and $\Lambda CDM$ model decrease with the increase of ${{\Lambda }_{0}}$ in the initial value $\mathscr K\left( {{t}_{0}} \right)={{H}_{0}}\left( \sqrt{1-\dfrac{\Lambda -{{\Lambda }_{0}}}{3{{H}_{0}}^{2}}}-1 \right)$  case, while the latter is a little bit larger than the former. The difference between {\bf Case C} and $\Lambda CDM$ model increases with the increase of $w_0$ and reaches to a quite large deviation when $w_0=-\dfrac{1}{3}$. 

The behavior of luminosity distance $d_L$ versus redshift $z$ curves for the initial value $\mathscr K\left( {{t}_{0}} \right)=-{{H}_{0}}\left( \sqrt{1-\frac{\Lambda -{{\Lambda }_{0}}}{3{{H}_{0}}^{2}}}\text{+}1 \right)$ case is similar to one of  $\mathscr K\left( {{t}_{0}} \right)={{H}_{0}}\left( \sqrt{1-\dfrac{\Lambda -{{\Lambda }_{0}}}{3{{H}_{0}}^{2}}}-1 \right)$  case. 

The distance modulus is defined as ${\mu=25+5\log_{10}\left(d_L/Mpc\right)}$\cite{nielsen2016marginal}.  
\begin{figure}[!htbp]

	\centering	
	\includegraphics[width=7in]{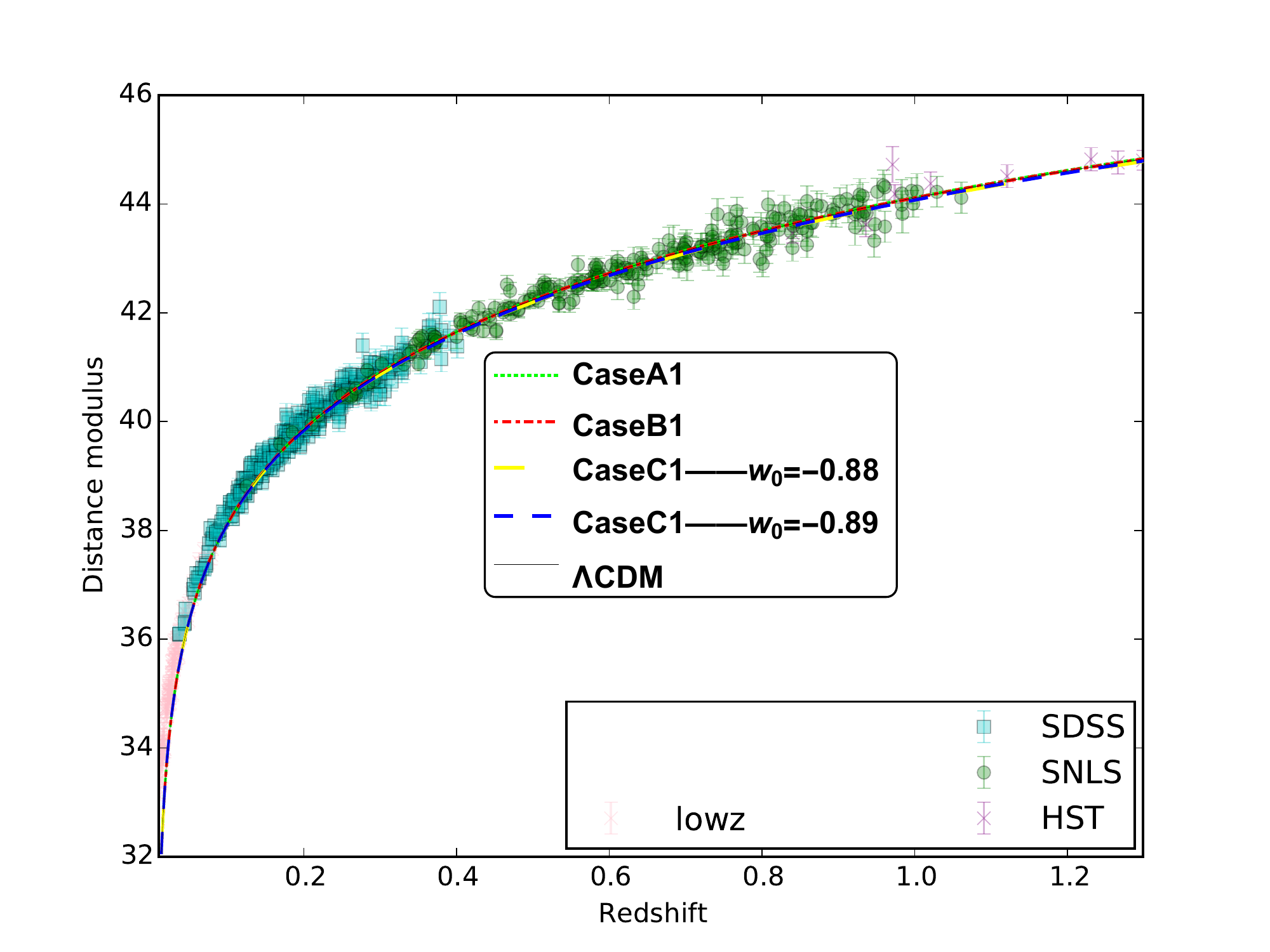}
	\caption{the comparison between theoretical prediction of the three case of approximation and observation on luminosity distance modulus in the case of initial value $\mathscr K\left( {{t}_{0}} \right)={{H}_{0}}\left( \sqrt{1-\frac{\Lambda -{{\Lambda }_{0}}}{3{{H}_{0}}^{2}}}-1 \right)$, in the case of $\Lambda_0=0.2\Lambda$ }\label{muzk0pl0_02l}
\end{figure}

\section{The Effective Quintessence Potential from String Landscape }

By the comparison of eq.\eqref{rhoK} and eq.\eqref{mfeq1}, one can introduce the idea of effective cosmological constant which is responsible for the accelerating expansion of the universe rather than the bare one $\Lambda_0$,
\begin{equation}\label{Leff}
{{\Lambda }_{eff}}\left( t \right)={{\Lambda }_{0}}-3\left(\mathscr  K{{\left( t \right)}^{2}}+2\mathscr K\left( t \right)\frac{\dot{a}\left( t \right)}{a\left( t \right)} \right)\; .
\end{equation}
The bare cosmological constant $\Lambda_0$ is just the vacuum energy density  which plays only a partial role in the accelerating expansion in our approach. It is the non-trivial contortion which determines the accelerating expansion eventually.

Phenomenologically, the $\Lambda_{eff}$ can be regarded as an energy density produced by some auxiliary fields which are responsible for the accelerating expansion such as quintessence field etc\cite{Tsujikawa:2013fta}. We can consider the action for gravity in the form of
\begin{equation}\label{quintessence}
S_q=\int d^4x\sqrt{-g}\left[\dfrac{1}{2}M_{pl}^2R-\dfrac{1}{2}g^{\mu\nu}\partial_{\mu}\phi\partial_{\nu}\phi-V(\phi)\right]\;  .
\end{equation} 
The corresponding energy density for the field $\phi$ is
\begin{equation}\label{rhophi}
\rho_{\phi}=\dfrac{{\dot \phi}^2}{2}+V(\phi)
\end{equation}
and the pressure of field $\phi$ is
\begin{equation}\label{pphi}
p_{\phi}=\dfrac{{\dot \phi}^2}{2}-V(\phi)
\end{equation}
The continuity equation for the $\phi$ field is
\begin{equation}\label{eomphi}
\ddot{\phi}+3H\dot{\phi}+V_{,\phi}=0
\end{equation}
The solution of eq.\eqref{eomphi} would give the evolution $\phi(t)$. 
In the analog of LSLV model by quintessence, we have 
\begin{equation}\label{leffv}
\Lambda_{eff}(t)=\dfrac{{\dot \phi}^2(t)}{2}+V(\phi(t))
\end{equation}
Then we have the relation
\begin{equation}\label{leffdot}
V\left( \phi \left( t \right) \right)={{\Lambda }_{eff}}+\frac{{{{\dot{\Lambda }}}_{eff}}}{6H}
\end{equation}
 

\begin{figure}[!htbp]

	\centering	
	\subfigure[]
	{\label{lcrtcasea1}
		\includegraphics[width=3in]{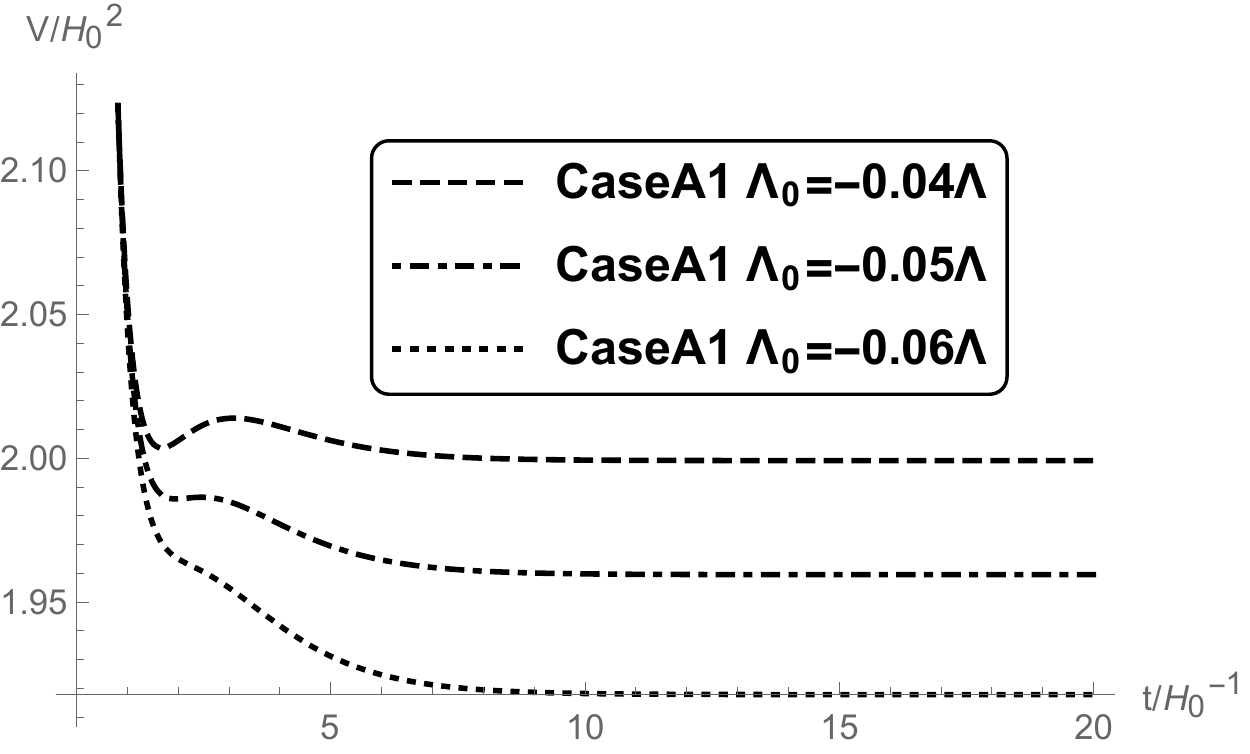}
	}
	\subfigure[]
	{\label{lcrtcaseb1}
		\includegraphics[width=3in]{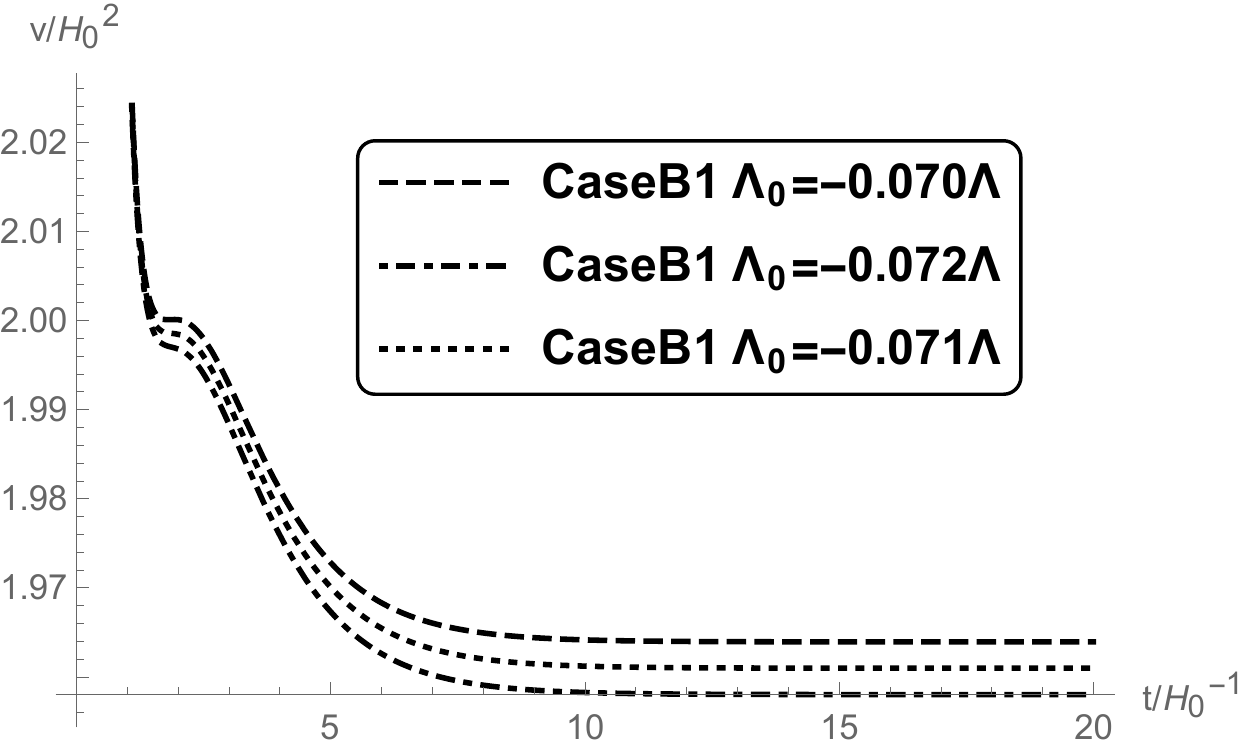}
	}
	\subfigure[]
	{\label{lcrtcasec1_w0m1}
		\includegraphics[width=3in]{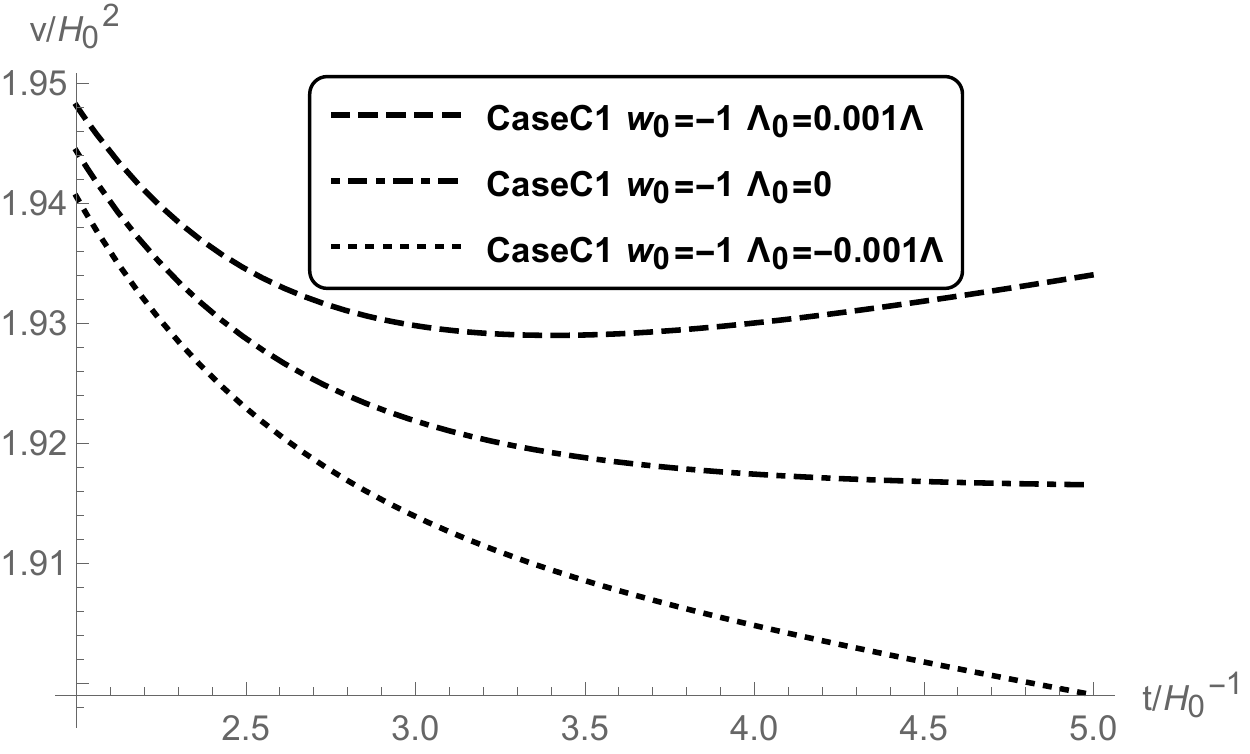}
	}
	\subfigure[]
	{\label{lcrtcasec1_w0m89}
		\includegraphics[width=3in]{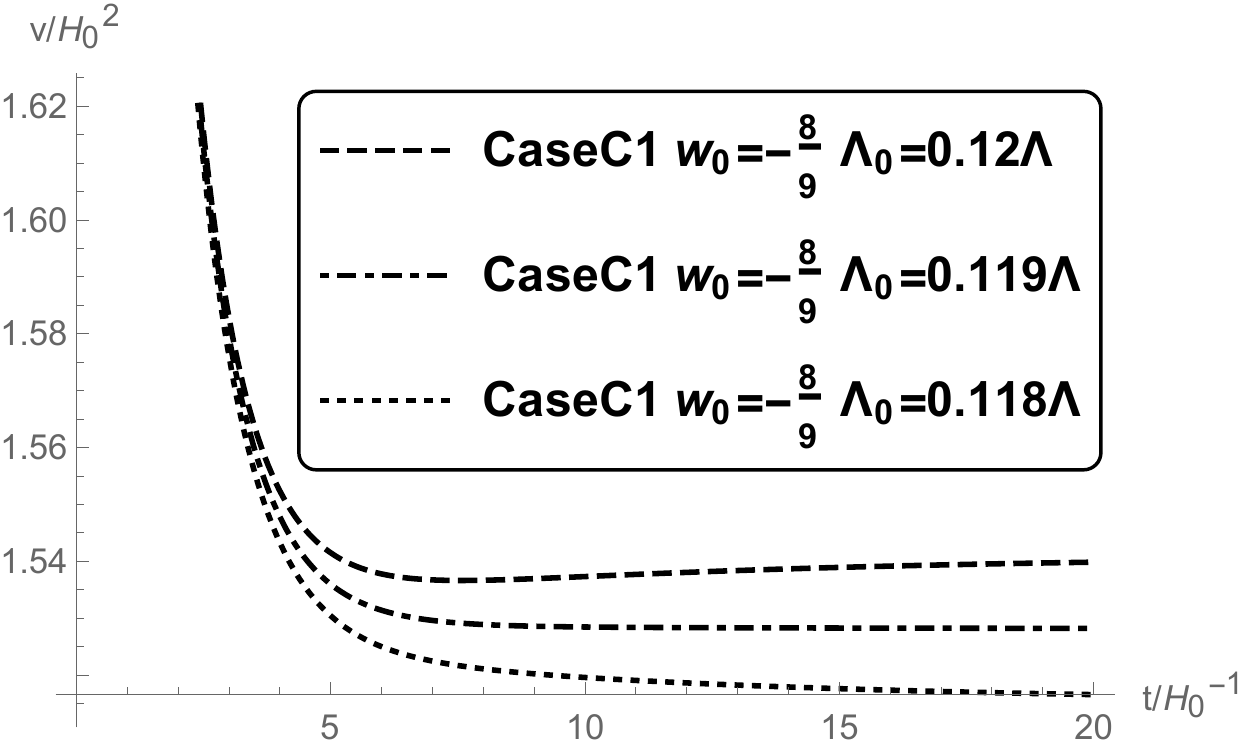}
	}
	\subfigure[]
	{\label{lcrtcasea2}
		\includegraphics[width=3in]{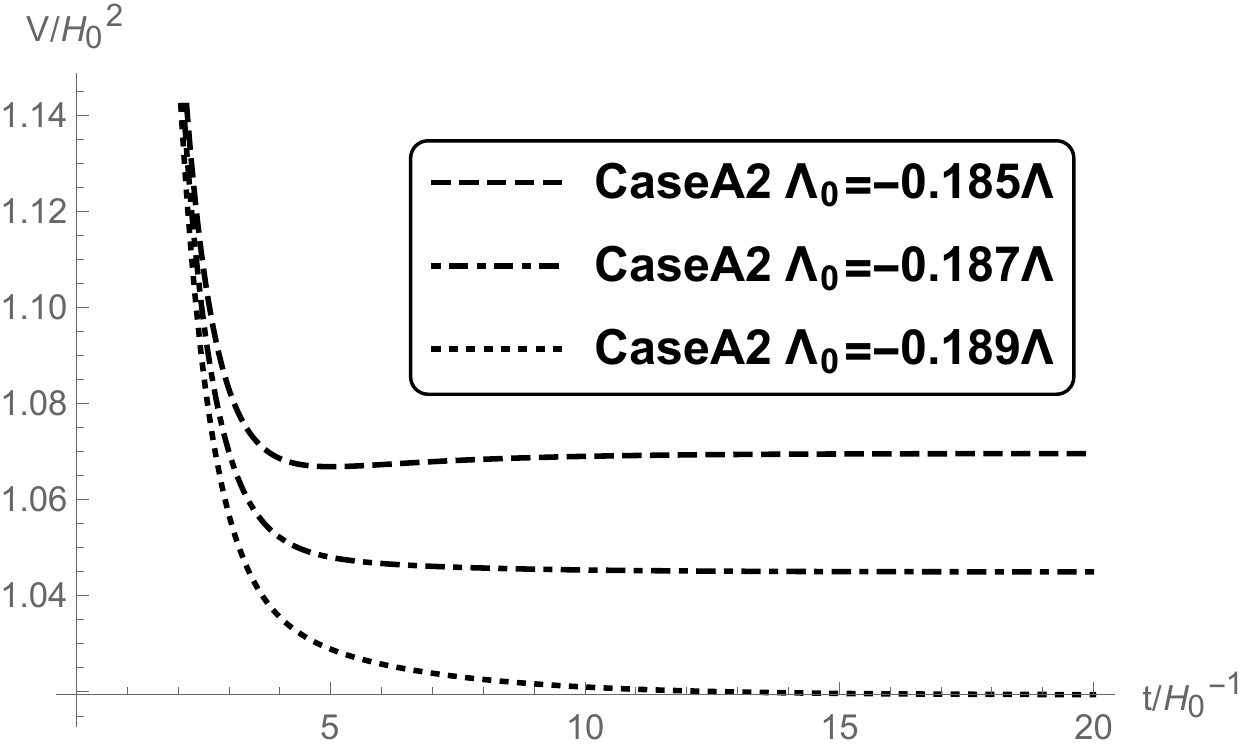}
	}
	\subfigure[]
	{\label{lcrtcaseb2}
		\includegraphics[width=3in]{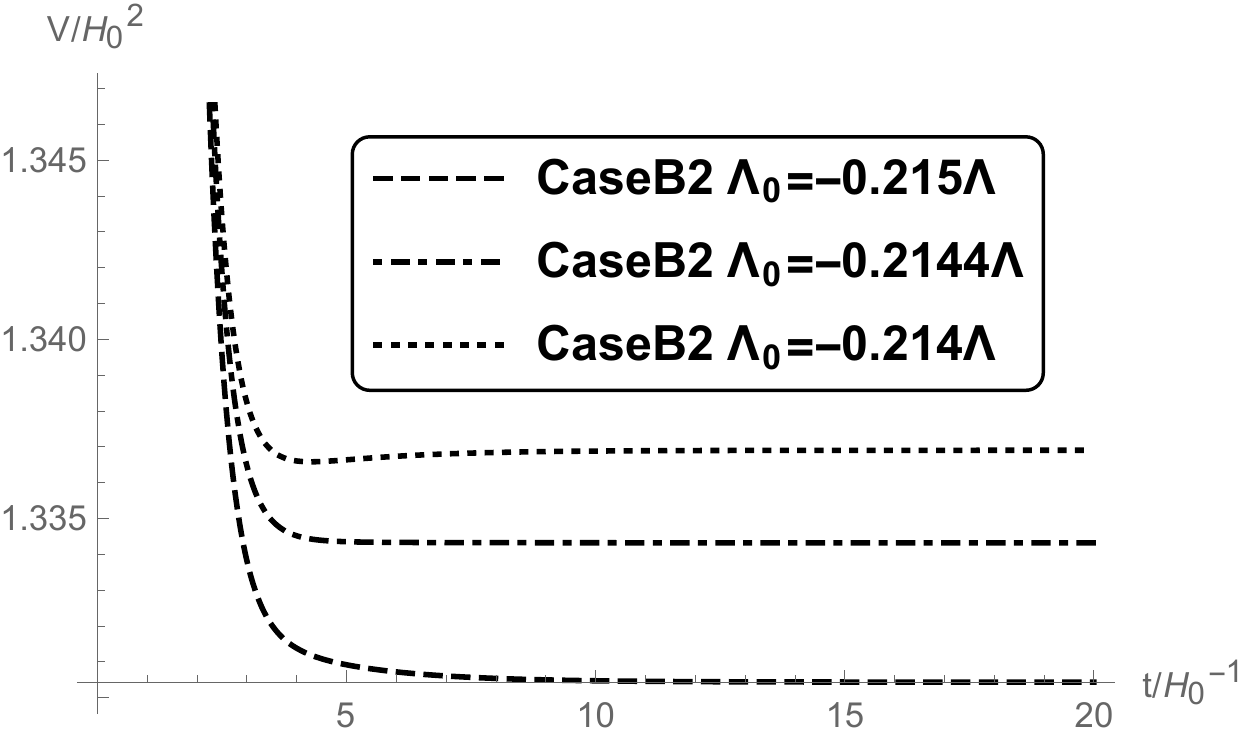}
	}
	\subfigure[]
	{\label{lcrtcasec2_w0m1}
		\includegraphics[width=3in]{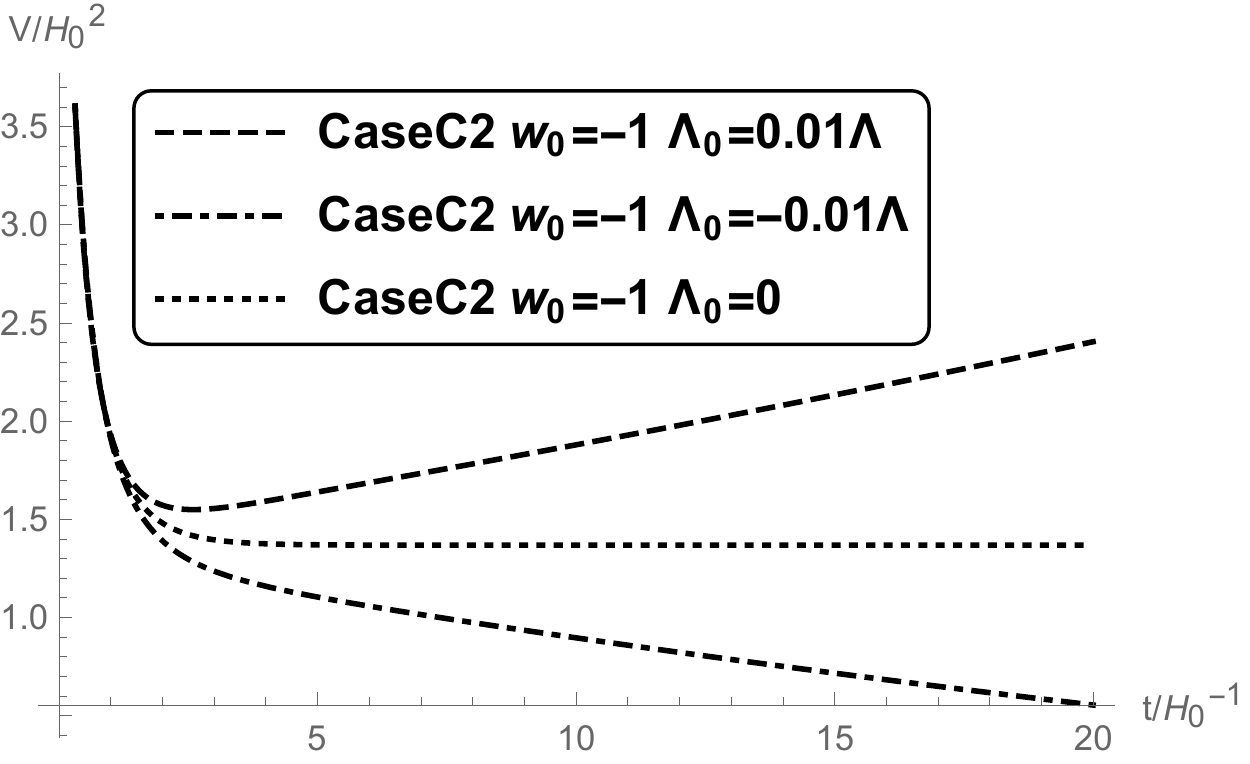}
	}
	\subfigure[]
	{\label{lcrtcasec2_w0m89}
		\includegraphics[width=3in]{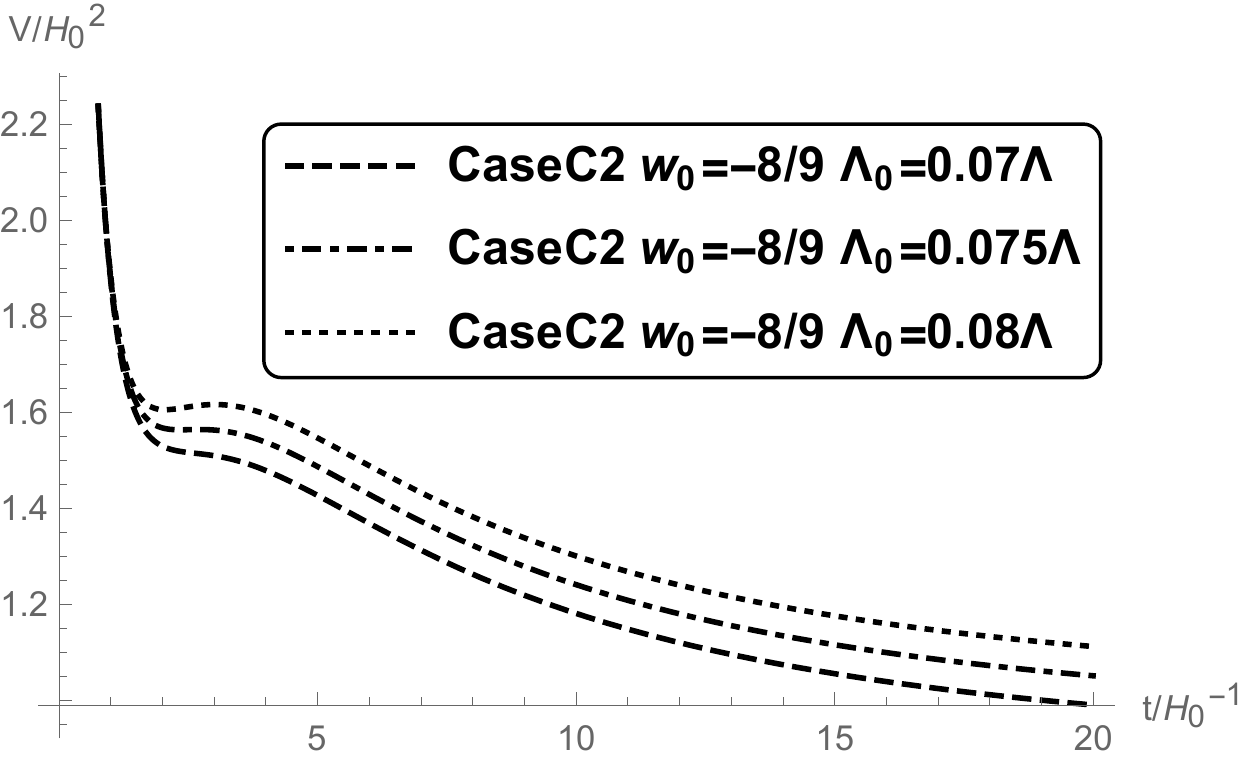}
	}
	\caption{the $\Lambda_{crit}$ solutions in all the cases with both the initial values $\mathscr K\left( {{t}_{0}} \right)={{H}_{0}}\left( \pm \sqrt{1-\frac{\Lambda -{{\Lambda }_{0}}}{3{{H}_{0}}^{2}}}-1 \right)$  }\label{lcrt}
\end{figure}

The evolution of $V\left( \phi \left( t \right) \right)$ given by $\Lambda_{eff}(t)$ versus $t$ bifurcates at some critical value of $\Lambda_0$, named as $\Lambda_{crit}$, i.e. $V\left( \phi \left( t \right) \right)$ decreases monotonically along with the increase of $t$ when $\Lambda_0 \le \Lambda_{crit}$ while its evolution has a local minimum when $\Lambda_0 > \Lambda_{crit}$. Table \ref{Lcrit} summarizes the value of $\Lambda_{crit}$ for all the cases of approximation in consideration. Fig.\ref{lcrt} shows the bifurcation of the evolution of $V\left( \phi \left( t \right) \right)$ versus $t$ with $\Lambda_0$ around $\Lambda_{crit}$ in all the cases with both the initial values of $\mathscr K (t)$.
We find that the {\bf Case C} in the $\mathscr K$'s first initial value case, $\mathscr K(t_0)={{H}_{0}}\left( \sqrt{1-\dfrac{\Lambda -{{\Lambda }_{0}}}{3{{H}_{0}}^{2}}}-1 \right)$, $V\left( \phi \left( t \right) \right)$ keeps decreasing monotonically along with the increase of $t$ for all value of $\Lambda_0$ allowed when $w_0 >-\dfrac{8}{9}$, i.e. the solution of $\Lambda_{crit}$ does not exist in this case. However, the comparison of theoretical prediction and observation on luminosity distance modulus reveals {\bf Case C1} when $w_0 >-\dfrac{8}{9}$ can be ruled out by observation on luminosity distance modulus versus redshift curve as is shown in Fig.\ref{muzk0pl0_02l}. It is reasonable to argue that $w_0 >-\dfrac{8}{9}$ for {\bf Case C1} is not a good approximation to fix the evolution of  $\mathscr K$.

\begin{table}[!htbp]
	\centering
	\caption{The critical values of $\Lambda_0$ triggers the transition from monotonic evolution of $V\left( \phi \left( t \right) \right)$ to the one with local minimum in different models}\label{Lcrit}
	\begin{tabular}{ccc}
		\toprule
		&Initial Values of $\mathscr K(t)$
		&Critical Values of $\Lambda_0$\\
		\midrule
		\multirow{2}{*}{{\bf Case A}}
		& $\mathscr K\left( {{t}_{0}} \right)={{H}_{0}}\left( \sqrt{1-\dfrac{\Lambda -{{\Lambda }_{0}}}{3{{H}_{0}}^{2}}}-1 \right)$
		&-0.05$\Lambda$ \\
		
		& $\mathscr K\left( {{t}_{0}} \right)=-{{H}_{0}}\left( \sqrt{1-\dfrac{\Lambda -{{\Lambda }_{0}}}{3{{H}_{0}}^{2}}}+1 \right)$
		& -0.187$\Lambda$\\
		\hline
		\multirow{2}{*}{{\bf Case B}}
		& $\mathscr K\left( {{t}_{0}} \right)={{H}_{0}}\left( \sqrt{1-\dfrac{\Lambda -{{\Lambda }_{0}}}{3{{H}_{0}}^{2}}}-1 \right)$
		&-0.071$\Lambda$ \\
		
		& $\mathscr K\left( {{t}_{0}} \right)=-{{H}_{0}}\left( \sqrt{1-\dfrac{\Lambda -{{\Lambda }_{0}}}{3{{H}_{0}}^{2}}}+1 \right)$
		& -0.2144$\Lambda$\\
		\hline
		\multirow{2}{*}{{\bf Case C($w_0=-1$)}}
		& $\mathscr K\left( {{t}_{0}} \right)={{H}_{0}}\left( \sqrt{1-\dfrac{\Lambda -{{\Lambda }_{0}}}{3{{H}_{0}}^{2}}}-1 \right)$
		&0.00001$\Lambda$ \\
		
		& $\mathscr K\left( {{t}_{0}} \right)=-{{H}_{0}}\left( \sqrt{1-\dfrac{\Lambda -{{\Lambda }_{0}}}{3{{H}_{0}}^{2}}}+1 \right)$
		& 0.00001$\Lambda$\\
		\hline
		\multirow{2}{*}{{\bf Case C($w_0=-\dfrac{8}{9}$)}}
		& $\mathscr K\left( {{t}_{0}} \right)={{H}_{0}}\left( \sqrt{1-\dfrac{\Lambda -{{\Lambda }_{0}}}{3{{H}_{0}}^{2}}}-1 \right)$
		&0.119$\Lambda$ \\
		
		& $\mathscr K\left( {{t}_{0}} \right)=-{{H}_{0}}\left( \sqrt{1-\dfrac{\Lambda -{{\Lambda }_{0}}}{3{{H}_{0}}^{2}}}+1 \right)$
		& 0.075$\Lambda$\\
		\hline
		\multirow{2}{*}{{\bf Case C($w_0=-\dfrac{7}{9}$)}}
		& $\mathscr K\left( {{t}_{0}} \right)={{H}_{0}}\left( \sqrt{1-\dfrac{\Lambda -{{\Lambda }_{0}}}{3{{H}_{0}}^{2}}}-1 \right)$
		&none \\
		
		& $\mathscr K\left( {{t}_{0}} \right)=-{{H}_{0}}\left( \sqrt{1-\dfrac{\Lambda -{{\Lambda }_{0}}}{3{{H}_{0}}^{2}}}+1 \right)$
		& 0.143$\Lambda$\\
		\hline
		\multirow{2}{*}{{\bf Case C($w_0=-\dfrac{2}{3}$)}}
		& $\mathscr K\left( {{t}_{0}} \right)={{H}_{0}}\left( \sqrt{1-\dfrac{\Lambda -{{\Lambda }_{0}}}{3{{H}_{0}}^{2}}}-1 \right)$
		&none\\
		
		& $\mathscr K\left( {{t}_{0}} \right)=-{{H}_{0}}\left( \sqrt{1-\dfrac{\Lambda -{{\Lambda }_{0}}}{3{{H}_{0}}^{2}}}+1 \right)$
		& 0.2$\Lambda$\\
		\hline
		\multirow{2}{*}{{\bf Case C($w_0=-\dfrac{1}{3}$)}}
		& $\mathscr K\left( {{t}_{0}} \right)={{H}_{0}}\left( \sqrt{1-\dfrac{\Lambda -{{\Lambda }_{0}}}{3{{H}_{0}}^{2}}}-1 \right)$
		&none \\
		
		& $\mathscr K\left( {{t}_{0}} \right)=-{{H}_{0}}\left( \sqrt{1-\dfrac{\Lambda -{{\Lambda }_{0}}}{3{{H}_{0}}^{2}}}+1 \right)$
		& 0.306$\Lambda$\\
		\hline
		
		\bottomrule
	\end{tabular}
\end{table}
It can be observed that the solution of $\Lambda_{crit}$ values are all around $\Lambda_0=0$, the division of string landscape and swampland. We can make a reasonable analysis about the deviation of the $\Lambda_{crit}$ values from $\Lambda_0=0$. All the $\Lambda_{crit}$ values in Table \ref{Lcrit} are obtained with the approximation on the evolution of $\mathscr K$, we can guess $\Lambda_{crit}$ would be exactly zero in a more elaborated model of the evolution of $\mathscr K$. The monotonic $V(\phi)$ on $\phi$ will result a monotonic $V\left( \phi \left( t \right) \right)$ versus $t$. On the other hand, the appearance of local minimum in $V\left( \phi \left( t \right) \right)$ corresponds to a non-monotonic $V(\phi)$ on $\phi$. It reveals that $\Lambda_{crit}$ may be the real division of string landscape and swampland. 

To account for the inflation and the accelerating expansion, the AdS vacua of string landscape need to be lifted to dS ones by some unnatural mechanisms such as the KKLT or the LVS construction\cite{Kachru:2003aw,Cicoli:2008va} etc. However, the two criteria on dS swampland conjecture rule out the meta-stable dS types vacua as one with which the effective theory with UV completion is built and leave only the possibility of quintessence like vacua. The result obtained in this paper reveals that large scale Lorentz violation can lift the AdS vacua to quintessence like ones effectively.

\section{Summary and Outlook}
The KKLT or the LVS construction on uplifting the AdS vacua to dS ones relies on carefully adding $\bar{D_3}$-branes into the compactification. The $\bar{D_3}$-branes tension in a sufficiently warped background, in the presence of quantum corrections, can be a small enough correction to lift the formerly AdS vacuum to positive cosmological constant, without destabilizing the minimum. However, there are also many cases with no-go theorems to uplift the AdS vacua to dS ones. Several results reveal no-go theorems on dS
vacua in string theory constructions, with restrictions on ingredients used in string theory, typically specific combinations of fluxes, D-branes, orientifolds, etc\cite{Brennan:2017rbf}. Moreover, the second criterion on swampland conjecture exclude the effective theory with meta-stable dS vacua as theories with UV completion. It seems that to account for the accelerating expansion and inflation with a simple $\Lambda CDM$ model with positive cosmological constant is difficult. If the positive cosmological constant comes from vacuum energy density of a vacuum, both the construction of such vacuum is fragile and it will belong to the swampland. However, the quintessence like potential can pass the restriction by the second criterion of the swampland conjecture, which will also contribute a decreasing positive vacuum energy density. 

Our result obtained in this paper shows another route to uplift the AdS vacua to an effective quintessence types vacuum energy which can satisfy the second criterion of the swampland conjecture and without adding extra ingredients such as fluxes or D-branes into the modification. The result only relies on the consideration of large scale Lorentz violation from the quantum gravity frozen by the inflation. Actually, the origin causing accelerating expansion is assorted to the effect of quantum gravity in this scenario. The construction has a UV completion in this sense. 

Our approach is also different from quintessence model which has a scalar mode as a part of the gravity at large scale. The quintessence scalar field is either elementary, which needs detection examination, or an effective theory of other elementary physics mechanisms, which needs clarification on its relation with fundamental physics. In our approach, the quintessence field is actually an effective description of the dark partner contribution by contortion which is the effect of large scale LV and can be traced back to the quantum gravity origin. 

We employ three kinds of approximation to fix the evolution and magnitude of $\mathscr K(t)$. However, a more fundamental approach should be given later. We would start with a specific model of quantum gravity and a inflation model to achieve a model with a stronger ability of prediction on the evolution and magnitude of $\mathscr K(t)$. The Lorentz violation from quantum gravity can be traced along with the inflation and the frozen large scale Lorentz violation can be predicted hence. The present approach relies on gauge principle with the local symmetry group as the proper subgroup of Lorentz group. Actually, the Lorentz violation can be other types rather than the totally breakdown of some symmetry generators. The Lorentz violation can be achieved with a boost transformation differing from Lorentz boost only. Some discussions assert that the quantum gravity effect can modify the Poincar\'e algebra into a deformed one, a Hopf algebra e.g. $\kappa$-Poincar\'e etc(see e.g.\cite{Smolin:2010xa}). The gauge principle is not applicable in these cases. How to introduce the Lorentz violation in long range effective gravity beyond gauge principle is under investigation.  

We strengthen the idea proposed in\cite{Shen:2018elj} which stresses on the possibility that the LV in quantum gravity may be frozen at scale beyond horizon by inflation and transformed into a large scale LV afterwards which may play an important role in the evolution of the universe at late time. Actually we know that the physics at much lower energy scale than Planck one obeys Lorentz symmetry exactly and general relativity describes gravitation phenomenon successfully at least up to astronomical scale. The standard model of particle physics and the gravitation theory of general relativity as well as some ideas beyond standard model such as the grand unification, SUSY etc., can supply our understanding about the physics during the normal expanding period of the universe except the late time accelerating expansion, which seems to need new physics to explain, although we understand little about how the Lorentz symmetry emerges from a quantum gravity theory with LV, which is also quite challenging and beyond the scope of the present paper. On the other hand, it is indeed necessary to characterize the transition from large scale LV to relatively small scale Lorentz invariance according to our framework. Actually, $\mathscr K(t)$, the non-trivial component of contortion, is the seeking parameter to characterize the magnitude of large scale LV. The theoretical prediction on its dependence with length scale and its evolution are highly non-trivial for they depend on the understanding of quantum gravity and details of inflationary mechanism. However, the parameter $\mathscr K(t)$ is actually observable. 
 
In the most approaches on the extension of general relativity with torsion, the torsion or the contortion can not propogate and can only be nontrivial in the region of matter source distribution with spin\cite{Hehl:1976kj}. The large scale spacetime must be torsion free and so is the case for the universe. The space-time felt by matter motion described by the left-hand side of eq.\eqref{G=T} is the Riemann-Cartan space-time which is determined not only by energy-momentum but also by spin as in eq.\eqref{effangularmomentum}. In our approach, though the torsion caused by frozen large scale Lorentz violation propagates neither, it distributes as the shadow of matter distribution and evolves along as the evolution of the universe. Matter moves in a Riemannian spacetime instead of Riemann-Cartan one and interacts with the distribution of torsion in the usual way of gravitation, i.e. the effective energy momentum tensor contributed by contortion distribution participates the determination of spacetime curvature and matter moves along the geodesic line and feels the contortion effect through the spacetime curvature. However, the spin of matter particle can interact with the torsion tensor directly in addition to the gravitational interaction\cite{Speziale:2018cvy,Trukhanova:2018gmh,Trukhanova:2017yyq}. The large scale long propagation of particle with spin may exhibit the deviation from geodesic line and causes the advance or delay arrival time variation with its energy. There are indeed some indication of such events for the gamma-ray bursts and neutrinos and the delay differs between gamma-ray and neutrino\cite{Zhang:2018otj}. A detailed investigation on the magnitude of the distorted propagation caused by the large scale contortion distribution for both the neutrino and light events based on our approach of frozen large scale Lorentz violation is performing. It is expected that the comparison of observation with prediction will tell us the variation of large scale LV magnitude versus distance scale.

\section*{Acknowledgment}
This work is supported by the National Natural Science Foundation of China, under Grant No. 11775080 and Grant No. 11865016.

\bibliographystyle{utcaps1}
\bibliography{References}

\end{document}